\documentclass[11pt,aps,showpacs,showkeys]{revtex4}
\usepackage{amsmath,amsthm,amssymb,epsfig,alltt}

\begin{document}

%%%%%%%%%%%%%%%%%%%%%%%%%%%%%%%%%%%%%%%%%%%%%%%%%%%%%%%%%%%%%%%%%%%%%%%%%%%%%%%%%%%%%%%
\def\a{\alpha}
\def\b{\beta}
\def\d{{\delta}}
\def\l{\lambda}
\def\e{\epsilon}
\def\p{\partial}
\def\m{\mu}
\def\n{\nu}
\def\t{\tau}
\def\th{\theta}
\def\s{\sigma}
\def\g{\gamma}
\def\o{\omega}
\def\r{\rho}
\def\z{\zeta}
\def\D{\Delta}
\def\half{\frac{1}{2}}
\def\hatt{{\hat t}}
\def\hatx{{\hat x}}
\def\hatp{{\hat p}}
\def\hatX{{\hat X}}
\def\hatY{{\hat Y}}
\def\hatP{{\hat P}}
\def\haty{{\hat y}}
\def\whatX{{\widehat{X}}}
\def\whata{{\widehat{\alpha}}}
\def\whatb{{\widehat{\beta}}}
\def\whatV{{\widehat{V}}}
\def\hatth{{\hat \theta}}
\def\hatta{{\hat \tau}}
\def\hatrh{{\hat \rho}}
\def\hatva{{\hat \varphi}}
\def\barx{{\bar x}}
\def\bary{{\bar y}}
\def\barz{{\bar z}}
\def\baro{{\bar \omega}}
\def\barpsi{{\bar \psi}}
\def\sp{\sigma^\prime}
\def\nn{\nonumber}
\def\cb{{\cal B}}
\def\2pap{2\pi\alpha^\prime}
\def\pap{\pi\alpha^\prime}
\def\wideA{\widehat{A}}
\def\wideF{\widehat{F}}
\def\beq{\begin{eqnarray}}
 \def\eeq{\end{eqnarray}}
 \def\4pap{4\pi\a^\prime}
 \def\op{\omega^\prime}
 \def\xp{{x^\prime}}
 \def\sp{{\s^\prime}}
 \def\ap{\a^\prime}
 \def\tp{{\t^\prime}}
 \def\zp{{z^\prime}}
  \def\spp{\s^{\prime\prime}}
 \def\xpp{x^{\prime\prime}}
 \def\xppp{x^{\prime\prime\prime}}
 \def\barxp{{\bar x}^\prime}
 \def\barxpp{{\bar x}^{\prime\prime}}
 \def\barxppp{{\bar x}^{\prime\prime\prime}}
 \def\zetap{{\zeta^\prime}}
 \def\barchi{{\bar \chi}}
 \def\baro{{\bar \omega}}
 \def\bpsi{{\bar \psi}}
 \def\barg{{\bar g}}
 \def\barz{{\bar z}}
 \def\bareta{{\bar \eta}}
 \def\ta{{\tilde \a}}
 \def\tb{{\tilde \b}}
 \def\tc{{\tilde c}}
 \def\tz{{\tilde z}}
 \def\tJ{{\tilde J}}
 \def\tpsi{\tilde{\psi}}
 \def\tal{{\tilde \alpha}}
 \def\tbe{{\tilde \beta}}
 \def\tga{{\tilde \gamma}}
 \def\tchi{{\tilde{\chi}}}
 \def\barth{{\bar \theta}}
 \def\bareta{{\bar \eta}}
 \def\barom{{\bar \omega}}
 \def\bole{{\boldsymbol \epsilon}}
 \def\bolth{{\boldsymbol \theta}}
 \def\bomega{{\boldsymbol \omega}}
 \def\bolmu{{\boldsymbol \mu}}
 \def\bolal{{\boldsymbol \alpha}}
 \def\bolbe{{\boldsymbol \beta}}
 \def\bolL{{\boldsymbol  L}}
 \def\bolX{{\boldsymbol X}}
 \def\bbk{{\boldsymbol k}}
 \def\boln{{\boldsymbol n}}
 \def\bols{{\boldsymbol s}}
 \def\bolS{{\boldsymbol S}}
 \def\bola{{\boldsymbol a}}
 \def\bolA{{\boldsymbol A}}
 \def\bolJ{{\boldsymbol J}}
 \def\tr{{\rm tr}}
 \def\bbP{{\mathbb P}}
 \def\bbA{{\boldsymbol A}}
 \def\bbp{{\boldsymbol p}}
 \def\mathP{{\mathbb P}}

%%%%%%%%%%%%%%%%%%%%%%%%%%%%%%%%%%%%%%%%%%%%%%%%%%%%%%%%%%%%%%%%%%%%%%%%%%%%%%%%%%%%%%%%
\setcounter{page}{1}
\title[]{Cubic Closed String Field Theory on a Double Layer}

\author{Taejin Lee}
\email{taejin@kangwon.ac.kr}
\affiliation{Department of Physics, Kangwon National University, Chuncheon 24341 Korea}

\date{\today }

\begin{abstract}
Various studies have attempted to extend the Witten's cubic open string field theory to the cubic closed string field theory, which may describe Einstein gravity in the low energy sector consistently. In the present study we propose a cubic closed string field theory, introducing a double layer to describe the closed string world-sheet as an extension of the open string world-sheet of the Witten's cubic open string. We mapped the closed string world-sheet onto the complex plane, of which the lower half plane is completely covered by the extended part of the string world-sheet. Using the Green's function on the complex plane, evaluated the Polyakov string path integral, from which we extracted the Neumann functions and the vertex operators. We obtained the one and two string vertices (identities) following this procedure and finally the cubic string vertex operator. It is notable that the obtained cubic string scattering amplitude coincide with the cubic graviton interaction of the Einstein gravity. Thus, the resultant cubic closed string field theory consistently describes the Einstein gravity.  We also show that the Kawai-Lewellen-Tye (KLT) relations of the first quantized string theory may be manifested in the cubic closed string field theory.

\end{abstract}

%11.25.-w 	Strings and branes
%11.25.Db   Properties of perturbation theory
%11.25.Sq   Nonperturbative techniques; string field theory
%11.25.Hf   Conformal field theory, algebraic structures



\maketitle

\setcounter{footnote}{0}

\newpage
\tableofcontents
\newpage

\section{Introduction}

Since the seminal paper of Witten \cite{Witten1986,Witten92p}, the covariant string field theory has become one of the main pillar in string theory study \cite{Giddings86,Cremmer86,Samuel86,Giddings1986nucl,GiddingsPLB1986,Grossjevicki87a,Grossjevicki87b,Leclair89a,Leclair89b}. It is the Becchi-Rouet-Stora-Tyutin (BRST) symmetry \cite{Siegel1984aPLB,Siegel1984bPLB,Siegel1985aPLB,Siegel1985bPLB,Siegel1986Nucl,Banks1986,Hata1986}  that enabled him to construct the relativistic gauge covariant interacting string field theory. However, since the cubic string field theory is defined on 
a two dimensional surface with conic singularity, it is difficult to evaluate scattering amplitudes of order higher than three. This difficulty may be dealt with as studies have shown that the Witten's cubic string field theory is continuously deformable \cite{TLeeJKPS2017,Lee2017d,TLee2017cov,Lai2018,Lee2019PLBfour}  to a covariant cubic string field theory in the proper-time gauge \cite{Lee1988Ann} where the world-sheet is planar. Recently, it has been reported that the Witten's cubic string field theory is deformable \cite{Matsunaga2019,Erler2021} to the conventional light-cone string field theory \cite{Mandelstam1973,Mandelstam1974,Kaku1974a,Kaku1974b} , where the world-sheet is planar, hence free of conic singularity. 

In our study, we extended the covariant cubic string field theory in the proper-time gauge for open string to the cubic closed string theory. We confirmed that the closed string theory reproduces three-graviton-scattering amplitude as in \cite{TLeeEPJ2018} and the 
four graviton scattering amplitude of the Einstein gravity as in \cite{Lee2019four}. Since Witten's cubic open string field theory is deformable to the cubic 
open string field theory in the proper-time gauge, it must be possible to extend the
Witten's cubic open string field theory to the closed string field theory, which is deformable to the closed string field theory in the proper-time gauge. Based on this, we
re-examined the world-sheets of Witten's open string field theory. 

Some efforts have been made towards extending Witten's cubic open string theory to 
BRST invariant cubic closed string theory. Lykken and Raby proposed a set of axioms to construct the gauge invariant action for the interacting closed string \cite{Lykken1986}. In addition, Witten's field theory of the open bosonic string may also contain closed strings because the closed string poles appear at the level of loop diagrams of open string \cite{GiddingsPLB1986,Witten1986nucl,Srednicki1987nucl,Strominger1987PRLClose,Zwiebach1992}. Despite all these endeavors the task to construct a BRST invariant cubic closed string field theory has remained an unsolved problem.  

The world-sheet of Witten's cubic open string field theory, which is a conical surface,
may be mapped on a disk by a well-defined conformal transformation: The spatial coordinate on the two dimensional world-sheet is restricted to $[0,\pi]$ and the world trajectories of open string end points form a unit circle. By extending the range of the spatial coordinates $\eta_r$, $r=1,2,3$ to $[-\pi, \pi]$, then the image of the string stretches out the unit disk and form a closed line, if a periodic condition is imposed; describing a closed string. A double layer may be introduced for local patch, so that the extended parts may be put on the second layer. When the 
unit disk is further mapped to complex $z$-plane, it becomes clearer. The unit disk can be mapped onto the upper half complex $z$-plane by a simple conformal transformation.  Applying the same conformal transformation, the extended part of the string world-sheet is now mapped onto the lower half complex $z$-plane so that the entire complex $z$-plane is covered by the world-sheet of three interacting closed strings. 

The rest of this paper is organized as follows: In Section 2 we give a brief review of Witten's cubic open string field theory and introduce a double layer to describe the extended parts of string world-sheet. Section 3 defines the Fock space 
representation of vertex operators and the Neumann functions. In Section ,4 we describe the construction of one string identity and two-string vertex operators for closed string. The corresponding overlapping delta functional are identified and the world-sheets of two closed strings are mapped onto the complex $z$-plane. In Section 5, we discuss how extend Witten's cubic open string theory to the BRST invariant cubic closed string theory. The cubic interaction for the three closed strings has been explicitly
constructed. In Section 6, we calculated the Neumann functions for three-closed-string vertex operator. We reproduced the scattering amplitude of three gravitons of the Einstein gravity, using the cubic closed string field theory. In Section 7, we give a brief conclusion and a few remarks on the future work. The Appendix section devotes details of Neumann functions of closed string
vertex operators.

\section{Closed String Theory on a Double Layer}

\begin{figure}[htbp]
\begin {center}
\epsfxsize=0.6\hsize

% Specify the picture file name to be included.
\epsfbox{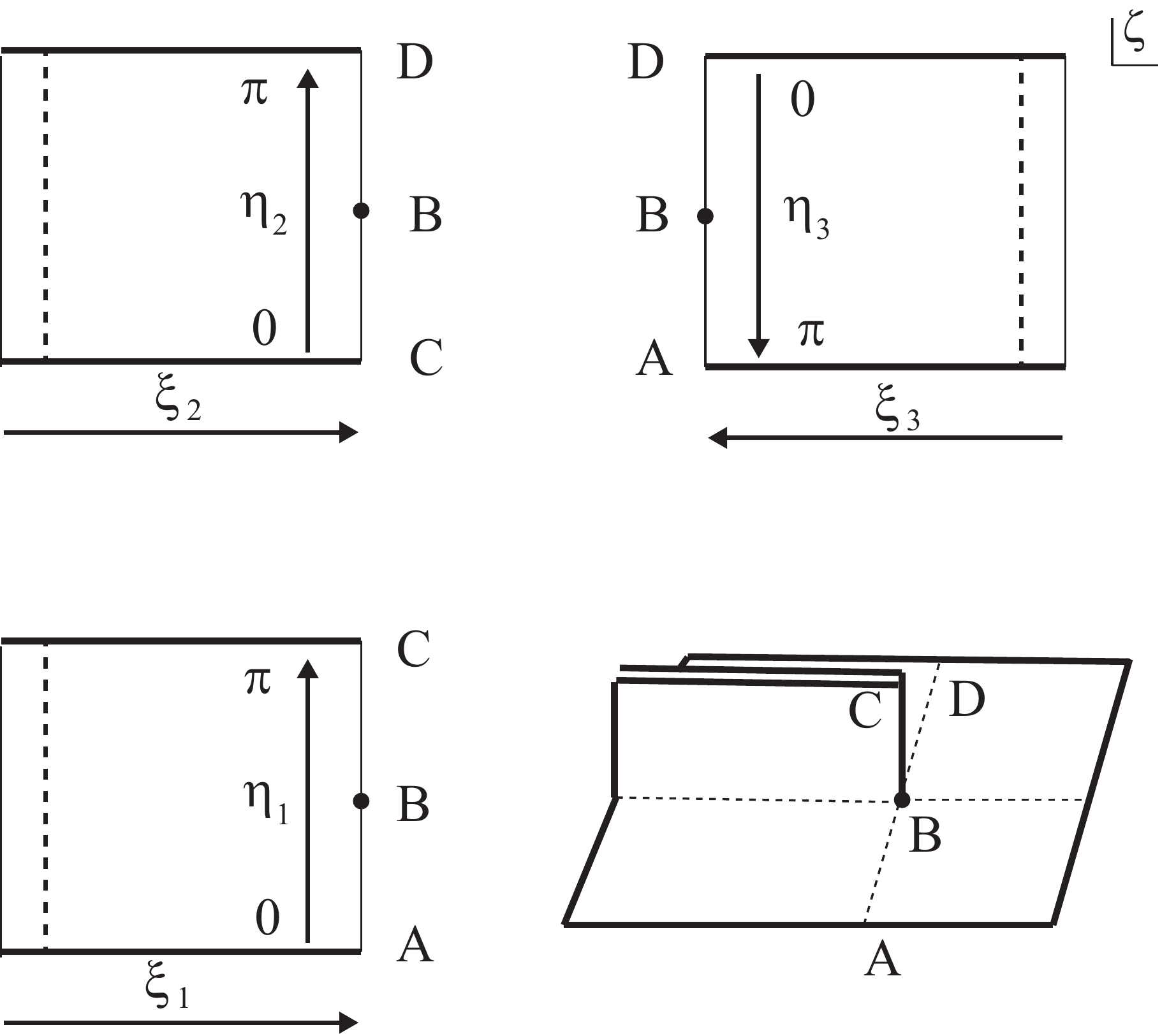}
\end {center}
\caption {\label{3patches} Local coordinate patches on the world-sheet of three-string scattering.}
\end{figure}

The Witten's cubic open 
string field theory is described by a BRST invariant action, which is given as 
\beq
S_{\rm open} &=& \int \text{tr} \left( \Psi * Q \Psi + \frac{2g}{3} \Psi * \Psi * \Psi \right),
\eeq
where the star product between the string field operators is defined as
\beq
\left(\Psi_1 * \Psi_2\right) [X(\s)] &=& \int \prod_{\frac{\pi}{2} \le \s \le \pi} DX^{(1)}(\s) \prod_{0 \le \s \le \frac{\pi}{2}} DX^{(2)}(\s)  \nn\\
&& 
\prod_{\frac{\pi}{2} \le \s \le \pi} \d \left[X^{(1)}(\s) - X^{(2)}(\pi -\s) \right] \Psi_1[X^{(1)}(\s)] \Psi_2[X^{(2)}(\s)]. \label{star}
\eeq 
The star product is associative and the string field action is invariant under the BRST gauge transformation
\beq \label{BRST}
\d \Psi = Q * \e + \Psi * \e - \e * \Psi .  
\eeq 
In terms of the normal modes, 
the open string coordinates $X^\m$, may be expanded as
\beq
X^\m (\s) &=& x^\m + 2 \sum_{n=1}^\infty \frac{1}{\sqrt{n}} x^\m_n \cos \left(n \s\right), ~~ 
0 \le  \s \le \pi, ~~ \m = 0, 1, \dots, d,
\eeq 
and the string field $\Psi$ may carry the group indices
\beq
\Psi[X]  =\frac{1}{\sqrt{2}} \Psi^0 [X] + \Psi^a[X] T^a,  ~~~ a =1, \dots, N^2-1 ,
\eeq
where $\Psi^0$ is the $U(1)$ component and $\Psi^a$, $a =1, \dots, N^2-1$ are the $SU(N)$ components. 
If we introduced three local coordinate patches, which describe free propagation of three open strings,
we might have depict the string world-sheet of three-open-string interaction as by Fig. \ref{3patches}.

%%%%%%%%%%%%%%%%%%%%%%%%%%%%%%%%%%%%%%%%%%%%%%%%%%%%%%%%%%%%%%%%%%%%%%%%%%%%%%%%%%%%%%%%%%
%%%%%%%%%%%%%%%%%%%%%%%%%%%%%%%%%%%%%%%%%%%%%%%%%%%%%%%%%%%%%%%%%%%%%%%%%%%%%%%%%%%%%%%%

The cubic string vertex operator, which is the Fock space representation of three open string interaction is obtained by mapping the string world-sheet onto the upper half complex plane,
where the Green's function is well-known. The mapping, called the Schwarz-Christoffel (SC) transformation, from the world-sheet onto the upper half plane is constructed in two steps: 
First, we map the world-sheet onto a unit disk by a conformal transformation:
%\begin{subequations}
\beq\label{cs1}
\omega_1 &=& e^{\frac{2\pi i}{3}} \left(\frac{1+ i e^{\zeta_1}}{1 - i e^{\zeta_1}}\right)^{\frac{2}{3}}, \nn\\
\omega_2 &=& \left(\frac{1+ i e^{\zeta_2}}{1 - i e^{\zeta_2}}\right)^{\frac{2}{3}},\\
\omega_3 &=& e^{-\frac{2\pi i}{3}} \left(\frac{1+ i e^{\zeta_3}}{1 - i e^{\zeta_3}}\right)^{\frac{2}{3}}. \nn
\eeq 
%\end{subequations}
Where the local coordinates on the three patches are given as $\zeta_r = \xi_r + i \eta_r$, $r= 1, 2, 3$.
Fig \ref{3patches}.  depicts three-open-string world-sheet mapped on a unit disk ($\o$-plane).   The interaction point $B$, where all the three open strings meet, is mapped to the origin of the disk and the external strings are located at 
$e^{\frac{2\pi i}{3}}, ~ 1, ~ e^{-\frac{2\pi i}{3}}$ respectively. 
Then, each local coordinate patch on the unit disk is mapped onto the 
upper half plane by the following conformal transformation:
\beq \label{cs2}
z = -i \,\frac{\omega_r -1}{\omega_r +1}, ~~~
\frac{\pi}{3} \le \arg\, \omega_r \le \frac{2\pi}{3}, ~~~ r =1, ~2,~ 3.
\eeq 
The three-open-string world-sheet mapped on the $z$-plane is described by Fig. \ref{3patches}. 
The external strings are mapped to three points on the real line
\beq
Z_1 = \sqrt{3}, ~~ Z_2 = 0, ~~~ Z_3 = - \sqrt{3}.
\eeq 
Having mapped the world-sheet of three strings onto the upper half plane, 
we can adopt the well-known Green's functions on the upper half plane,
\beq
G_N(z,\zp) &=& \ln \vert z- \zp \vert + \ln \vert z- z^{\prime *} \vert, ~~~\text{for} ~~\text{Neumann boundary condition}.
\eeq

The main objective of  the present work is to extend this cubic BRST invariant action of open 
string to that of the closed string, and to construct the three-closed-string vertex operator : The closed string coordinates $X$ are decomposed  into left-movers and right-movers
\beq
X(\t,\s) = X_L(\t+\s) + X_R (\t-\s) ,~~~ - \pi \le \s \le  \pi 
\eeq  
whose normal mode expansions are given as 
\begin{subequations}
\beq
X_L (\t,\s) &=& x_L + \sqrt{\frac{\ap}{2}} \, p_L(\t+\s) + i \sqrt{\frac{\ap}{2}} \,\sum_{n\not=0} \frac{1}{n}
\a_n e^{-in(\t+\s)}, \\
X_R (\t,\s) &=& x_R + \sqrt{\frac{\ap}{2}} \, p_R(\t-\s) + i \sqrt{\frac{\ap}{2}} \,\sum_{n\not=0} \frac{1}{n}
\ta_n e^{-in(\t-\s)}, 
\eeq 
\end{subequations}
where $x= x_L+ x_R$.  

\begin{figure}[htbp]
\begin {center}
\epsfxsize=0.4\hsize
%
% Specify the picture file name to be included.
\epsfbox{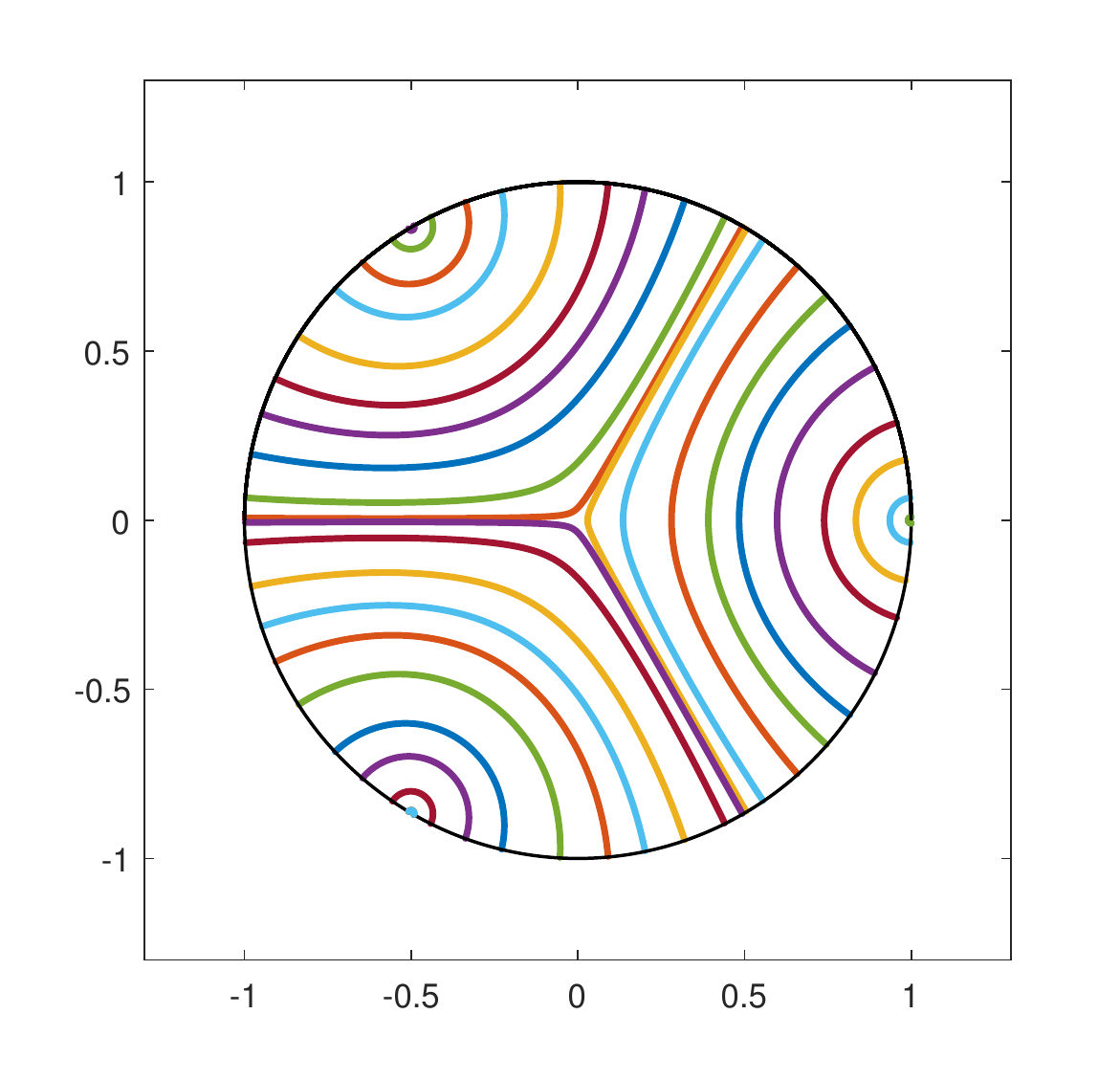}
\end {center}
\caption {\label{staropenw} Interaction of the three open strings on $\o$-complex plane.}
\end{figure}

\begin{figure}[htbp]
\begin {center}
\epsfxsize=0.4\hsize
%
% Specify the picture file name to be included.
\epsfbox{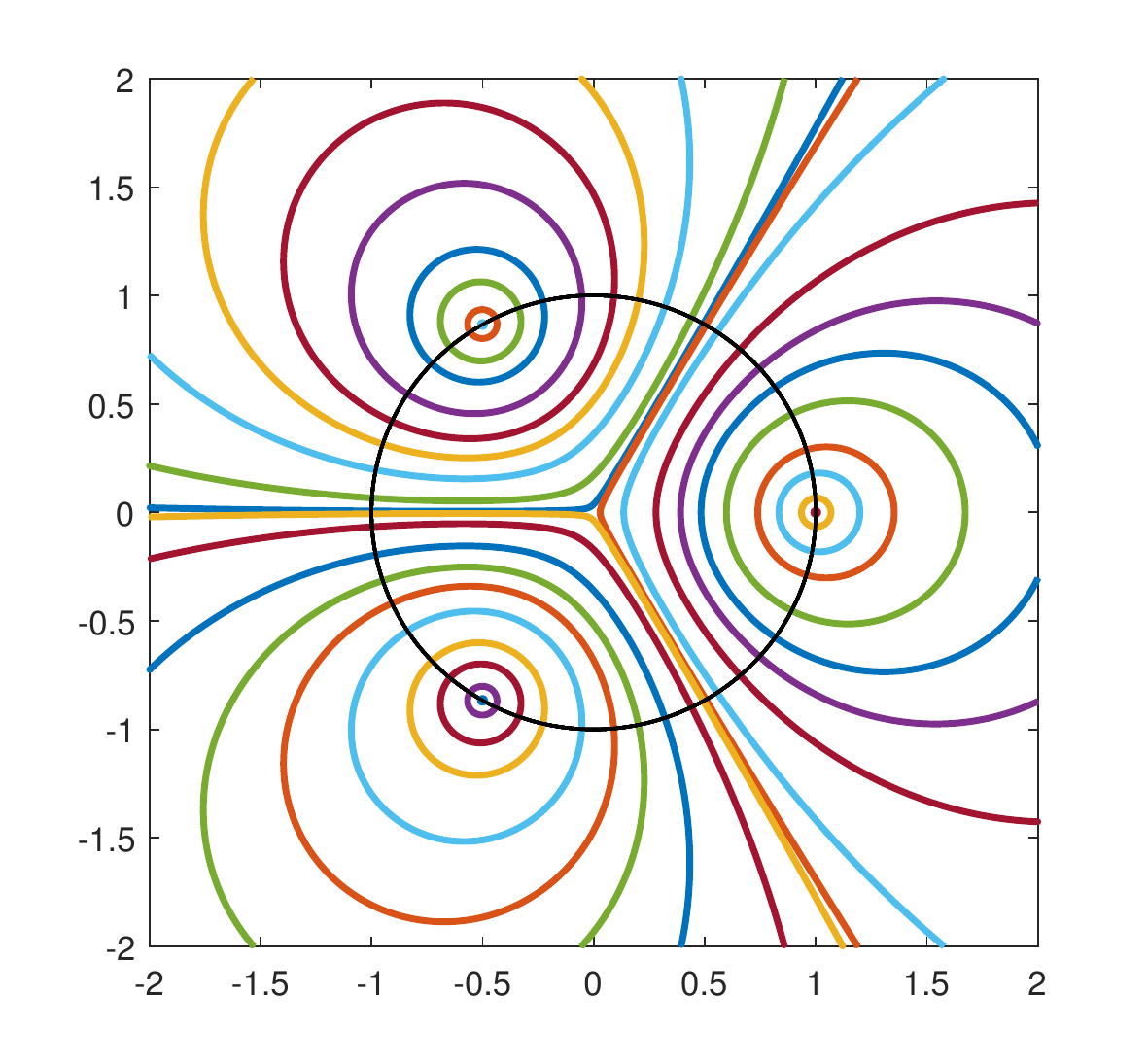}
\end {center}
\caption {\label{starclosedw} World-sheet of interacting three closed strings on $\o$-complex plane.}
\end{figure}

\begin{figure}[htbp]
\begin {center}
\epsfxsize=0.7\hsize
%
% Specify the picture file name to be included.
\epsfbox{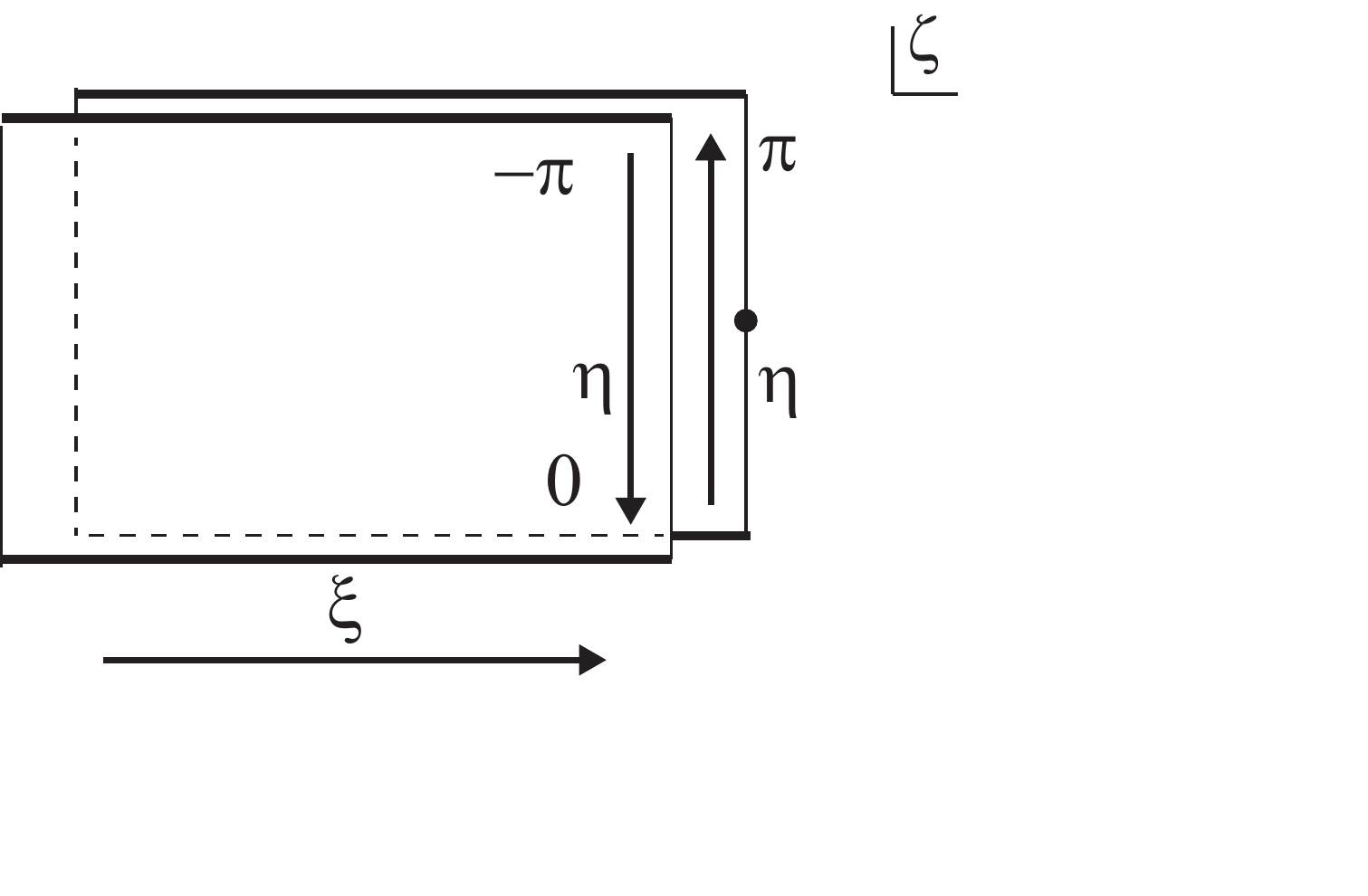}
\end {center}
\caption {\label{doub} String world-sheet with a Double Layer}
\end{figure}

Fig. \ref{staropenw} depicts world-sheet of cubic open string interaction where 
the ranges of $\eta_r$, $r=1, 2, 3$ are limited to $[0,\pi]$ . To describe a closed string, we may extend their 
ranges to $[-\pi, \pi]$,
\beq
|\o_r| &=& \left\vert \frac{1 + e^{2\xi_r} -2 e^{\xi_r}\sin \eta_r}{1 + e^{2\xi_r} +2 e^{\xi_r} \sin \eta_r} \right\vert^{\frac{1}{3}},~~r = 1, 2, 3.
\eeq 
By extending the ranges, the images of the extended parts of string, (for $-\pi \le \eta_r \le 0$) stretch out of the unit disk and make closed curves. (See Fig \ref{starclosedw}.)  Further, if we map the extended parts of the world-sheet onto the complex plane, they precisely fill the lower half complex plane. Therefore, it may be possible to describe the 
cubic closed string interaction by simply extending the ranges of $\eta_r$ with a periodic boundary condition. We may imagine that the string world-sheet has a double layer and the extended parts are on the second layer as depicted in Fig. \ref{doub} . We explored this possibility to study the cubic closed string throughout this work.

%%%%%%%%%%%%%%%%%%%%%%%%%%%%%%%%%%%%%%%%%%%%%%%%%
%%%%%%%%%%%%%%%%%%%%%%%%%%%%%%%%%%%%%%%%%%%%%%%%%

%%%%%%%%%%%%%%%%%%%%%%%%%%%%%%%%%%%%%%%%%%%%%%%%%
%%%%%%%%%%%%%%%%%%%%%%%%%%%%%%%%%%%%%%%%%%%%%%%%%

%\newpage
%%%%%%%%%%%%%%%%%%%%%%%%%%%%%%%%%%%%%%%%%%%%%%%%%
%%%%%%%%%%%%%%%%%%%%%%%%%%%%%%%%%%%%%%%%%%%%%%%%%

\section{The Neumann Functions for Closed String}

It is important to extract the Neumann functions from the cubic string interaction when studying
the string theory in terms of the component fields to compare it with the previously known field theory; in the case of closed string, the Einstein gravity. We obtained the overlapping delta functional for the closed string to define the cubic interaction in the (string coordinates) configuration space. We directly converted the overlapping functionals into the corresponding Fock space representations to get the Neumann functions, however, it was often difficult and may lead us to incorrect results. Throughout the present work worked with 
the Polyakov string path integral \cite{Polyakov1981} to extract the Neumann functions and the string vertex operators. 

Once we map the closed string world-sheet onto the complex plane, we adopt the  Green's function,
\beq\label{green}
G(z, \zp) &=&\langle X(\t,\s) X(\t^\prime,\s^\prime) \rangle = \ln \vert z - \zp \vert . 
\eeq 
The scattering amplitude of three strings, expressed as the Polyakov 
string path integral on the world-sheet with temporal boundary conditions fixed by the momenta of external strings \cite{GreenSW}:
\begin{subequations}
\beq
{\cal A} &=& \int \prod_{r, n} dP_{r,n} \Psi_r[P_r] W(P_{r,n}, p_r),  \\
W(P_{r,n}, p_r) &=& \int D[X] \exp \left(i \sum_r \int P^\m_r (\eta_r) X_\m (\xi_r, \eta_r) d\eta_r -\sum_r \int_{W_r} {\cal L} \, d\xi_r d\eta_r \right), \\
{\cal L} &=& \frac{1}{2\pi} \left\{ \left(\frac{\p X^\m}{\p \xi_r}\right)^2 + \left(\frac{\p X^\m}{\p \eta_r}\right)^2 \right\}.
\eeq 
\end{subequations} 
The wavefunction for the external strings are products of the oscillator eigenfunctions for each mode. 
The wave function for a state with occupation numbers $k^\m_{r,n}$ is given by 
\beq
\Psi\left(P^A_{r,n}\right) &=& \prod_{n=1} \prod_{A=1}^d H_{k^A_{r,n}}\left(P^A_{r,n}\right)
\exp \left[-\left(P^A_{r,n}\right)^2/4n \right] 
\eeq
where $H_{k^A_{r,n}}$ is the Hermite polynomial of degree $k^A_{r,n}$ in $P^A_{r,n}$. 
Using he Green's functions, Eq. (\ref{green}) and the Schwarz-Christoffel mapping, Eq. (\ref{cs1}). and Eq. (\ref{cs2}) , we may write 
$W(P_{r,n}, p_r)$ as follows
\beq
W(P_{r,n}, p_r) &=& g^{N-2+2L} \left[\det \Delta \right]^{-d/2} \exp \Biggr\{
\frac{1}{4} \sum_{r, s} \int d\sp d\spp \Bigl(P^{(r)}_\m(\sp) G(\sp, \t_r; \spp, \t_s) \eta^{\m\n} P^{(s)}_\n(\spp) \Bigr)\Biggl\}.
\eeq 
Evaluating the Polyakov string path integral and 
recasting it into the Feynman-Schwinger proper-time representation,
\beq
W(P_{r,n}, p_r)
&=&\langle  {\bf P} \vert \exp\left( \sum_r \xi_r L^{(r)}_0 \right) \vert V[3] \rangle 
\eeq 
where $\xi_r$, $r=1, \dots, N$ are the proper-times on local patches,
in the oscillatory basis, we obtain the Fock space representation of the $N$-closed-string-vertex 
$\vert V[N]\rangle $. 

The closed string Neumann functions are defined as Fourier components of the Green's function on complex plane 
\beq \label{greenclosed}
G(\r_r, \rho^\prime_s) &=&  \ln \vert z_r - z^\prime_s \vert \nn\\
&=& - \d_{rs} \left\{\sum_{n=1} \frac{1}{2n} \left(\omega^{-n}_+ \omega^{\prime n}_- 
+ \omega^{*-n}_+ \omega^{\prime * n}_-\right) - \max(\xi, \xi^\prime)  \right\} \nn\\
&&+ \sum_{n, m} \bar C^{rs}_{nm} e^{|n| \xi_r + |m| \xi^\prime_s} e^{in\eta_r} e^{im\eta^\prime_s}
\eeq
where $\rho_r$ and $\rho^\prime_s$ lie in the regions of the $r$-th and $s$-th string patches respectively and 
\beq
\o_r &=& e^{\zeta_r} = e^{\xi_r+ i\eta_r}, ~~~\op_s = e^{\xi^\prime_r + i\eta^{\prime}_s}, \\
(\o_+,\o_-) &=&  \left\{ 
\begin{array}{ll}
(\o_r, \op_s) , & ~~\mbox{for} ~~\xi_r \ge \xi^\prime_s   \\ 
(\op_s, \o_r) , & ~~\mbox{for} ~~\xi_r \le \xi^\prime_s 
\end{array} \right. \nn
\eeq

\section{Construction of Identity and Two-String Vertex for Closed String}

\subsection{One-Closed-String Vertex} 

\begin{figure}[htbp]
\begin {center}
\epsfxsize=0.4\hsize
%
% Specify the picture file name to be included.
\epsfbox{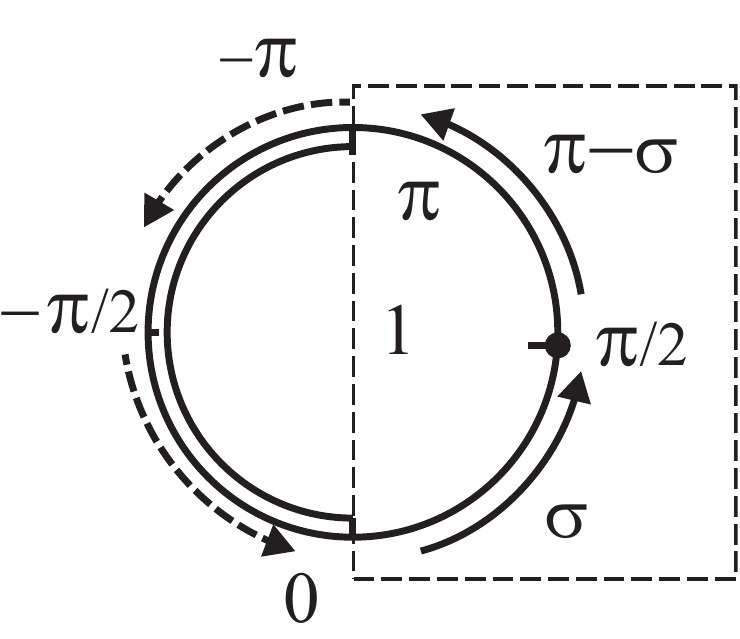}
\end {center}
\caption {\label{one} A Pictorial representation of the identity functional ($\s$-space).}
\end{figure}

The identity functional $I$ for the closed string with respect to $*$ may be given by an overlapping delta functional as 
\beq
I[X(\s)] = \langle X(\s) \vert I \rangle = \prod_{\frac{\pi}{2} \le \s \le \pi} \prod_{-\pi \le \s \le -\frac{\pi}{2} } \d \left(X(\s) - X(\pi -\s) \right)
\eeq 
It defines the one-string vertex operator. A pictorial representation of 
the overlapping delta functional is presented in Fig. \ref{one}.

%%%%%%%%%%%%%%%%%%%%%%%%%%%%%%%%%%%%%%%%%%%%%%%%%
%%%%%%%%%%%%%%%%%%%%%%%%%%%%%%%%%%%%%%%%%%%%%%%%%

%\subsubsection{External String is Located at $Z=0$} 

\begin{figure}[htbp]
\begin {center}
\epsfxsize=0.4\hsize
%
% Specify the picture file name to be included.
\epsfbox{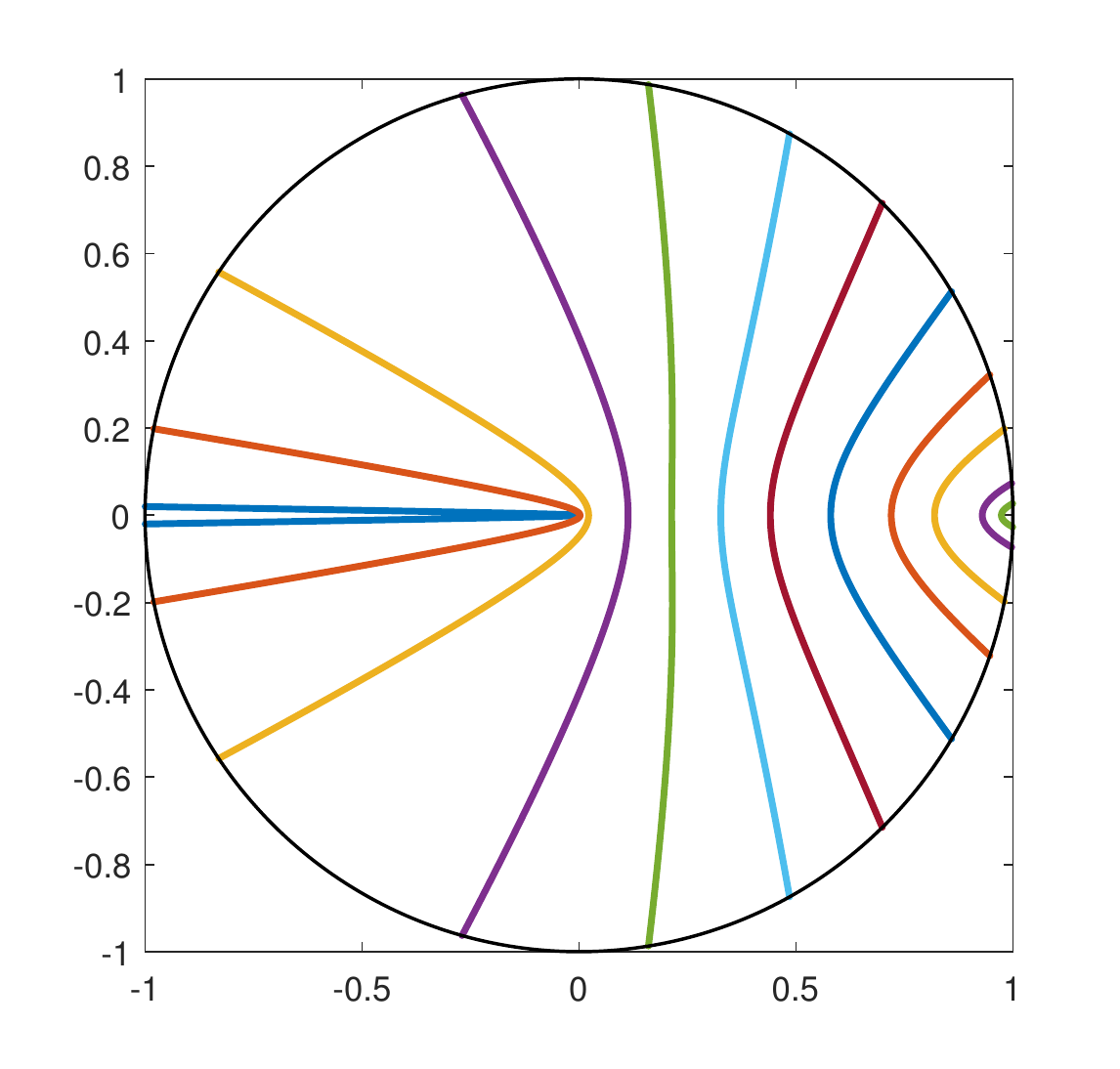}
\end {center}
\caption {\label{onewplane} One-open-string identity mapped onto a unit disk on the complex $\o$-plane.}
\end{figure}

If the external string was to be mapped onto $Z=0$ on the $z$-complex plane, we would choose the SC mapping from the string world-sheet to a unit disk  on the $\o$-plane as 
\beq
\omega &=& \left(\frac{1+ i e^{\zeta}}{1 - i e^{\zeta}}\right)^2, ~~~ -\pi \le \eta \le \pi .
\eeq
(See Fig.  \ref{onewplane} )
%%%%%%%%%%%%%%%%%%%%%%%%%%%%%%%%%%%%%%%%%%%%%%%%%
%%%%%%%%%%%%%%%%%%%%%%%%%%%%%%%%%%%%%%%%%%%%%%%%%

\begin{figure}[htbp]
\begin {center}
\epsfxsize=0.5\hsize
%
% Specify the picture file name to be included.
\epsfbox{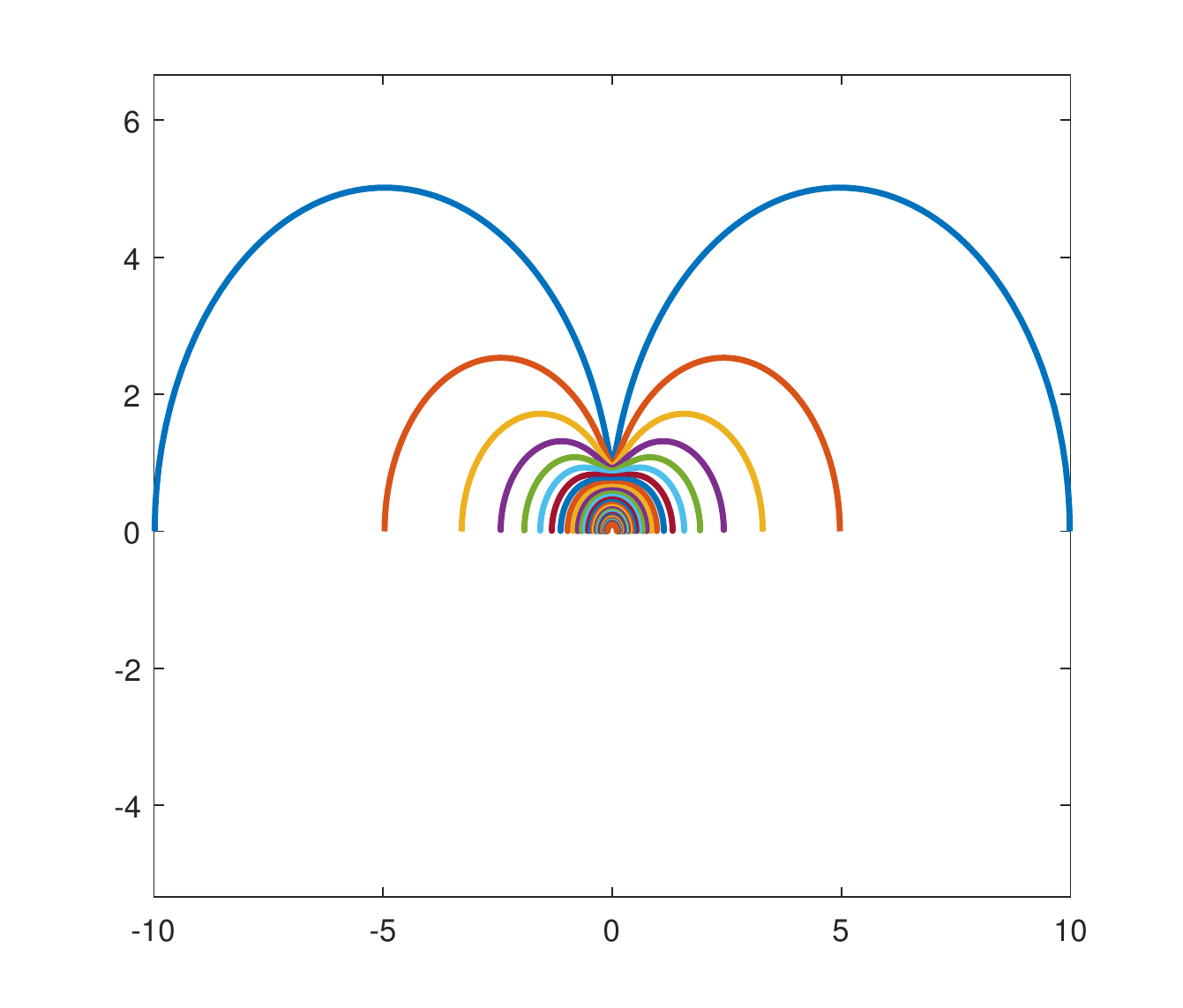}
\end {center}
\caption {\label{oneopenz} One-open-string identity on complex $z$-plane.}
\end{figure}

\begin{figure}[htbp]
\begin {center}
\epsfxsize=0.5\hsize
%
% Specify the picture file name to be included.
\epsfbox{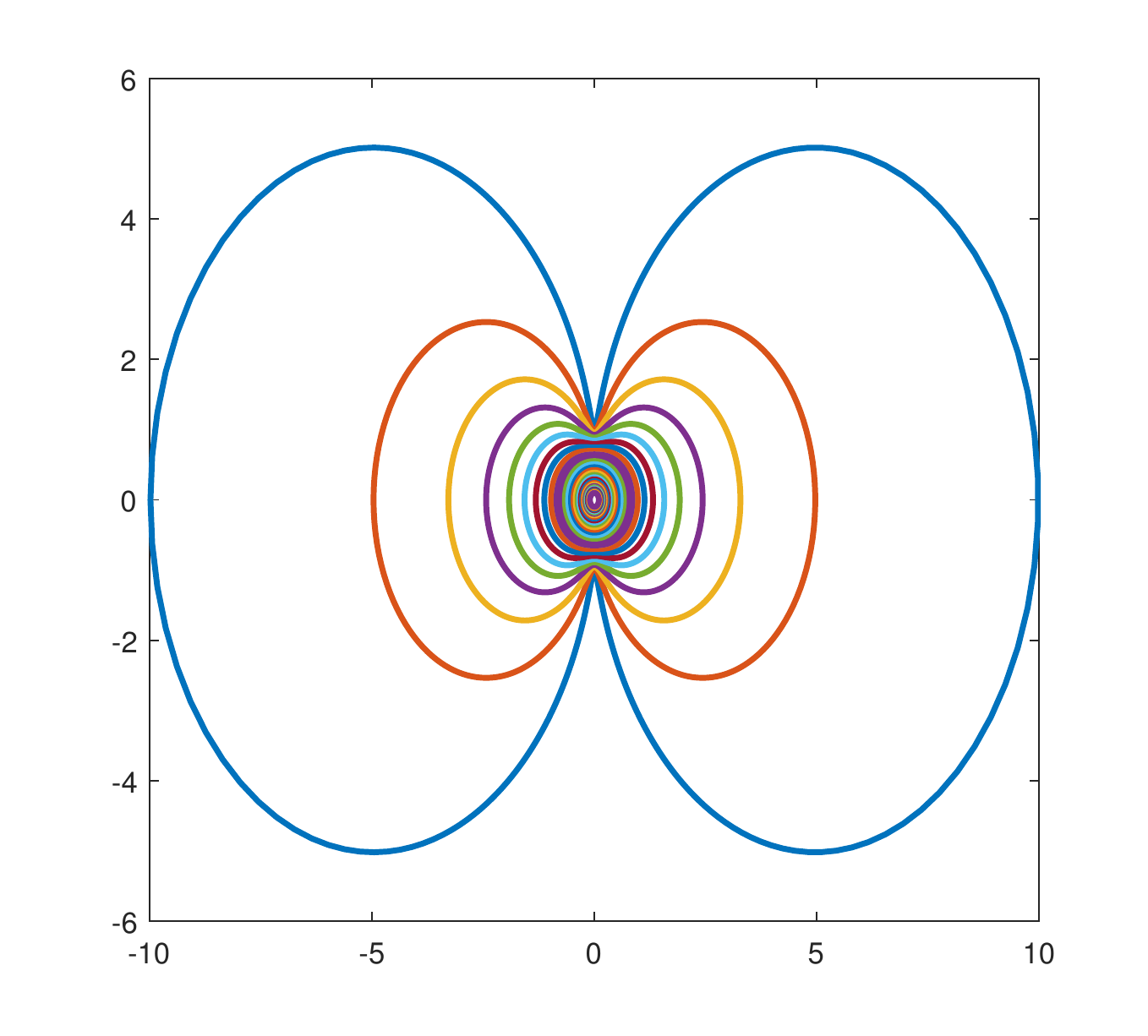}
\end {center}
\caption {\label{oneclosedz} One-closed-string identity on complex $z$-plane.}
\end{figure}

\noindent 
On the $\omega$-plane, the external string is located at $\o=1$. 
The disk can be mapped onto the $z$-complex plane by a well-known conformal mapping
\beq
z = -i \frac{\o -1}{\o + 1}=-i \frac{\left(\frac{1+ i e^{\zeta}}{1 - i e^{\zeta}}\right)^2 -1}{\left(\frac{1+ i e^{\zeta}}{1 - i e^{\zeta}}\right)^2 + 1} .
%, ~~~~~\o = \left(\frac{1 +iz}{1-iz} \right)
\eeq
If we restrict the range of $\eta$ to $[0,\pi]$ to describe the open string identity, the world-sheet of the unit disk is mapped onto the upper half of the complex $z$-plane (See Fig. \ref{oneopenz}).  
Since 
\beq
\o &=& \frac{1-2 e^{2 \xi} + e^{4 \xi} -4 e^{2 \xi} \cos^2 \eta + 4i e^{\xi} (1- e^{2\xi}) \cos \eta}{\left(
1 + e^{2\xi} + 2 e^\xi \sin^2\eta\right)^2}
\eeq 
and  $\o(\xi, \eta) = \o(\xi, -\eta)$, if we extend the range of $\eta$ to $-\pi \le \eta \le \pi$, the string world trajectory still remains inside of the unit disk, $\vert \o \vert \le 1$. 
However, the extension makes difference on the complex $z$-plane: If we limit the range of $\eta$ to $[0,\pi]$, corresponding to open string, the image of the string world-sheet 
is mapped onto the upper half complex $z$-plane as shown in 
Fig. \ref{oneopenz} . 
However, the extended part where $\eta \in [-\pi,0]$, covers the lower half of the complex $z$-plane. Thus, the entire $z$-complex plane is covered if we extend the range of $\eta \in [-\pi,\pi]$. (See Fig. \ref{oneclosedz})

%%%%%%%%%%%%%%%%%%%%%%%%%%%%%%%%%%%%%%%%%%%%%%%%%
%%%%%%%%%%%%%%%%%%%%%%%%%%%%%%%%%%%%%%%%%%%%%%%%%

%%%%%%%%%%%%%%%%%%%%%%%%%%%%%%%%%%%%%%%%%%%%%%%%%
%%%%%%%%%%%%%%%%%%%%%%%%%%%%%%%%%%%%%%%%%%%%%%%%%

The Fock space representation of the one-closed-string identity (vertex) is obtained from expansion of $e^{-\z}$ around $Z=0$
\beq
e^{-\z} &=& - i \left(\frac{1+ \o^\half }{1-\o^\half} \right) \nn\\
&=& \frac{2}{z}+\frac{z}{2}-\frac{ z^3}{8}+\frac{z^5}{16}-\frac{5 z^7}{128}+\frac{7 z^9}{256}+O\left(z^{11}\right) \nn\\
&=& \frac{2}{z} + \sum_{n=0} c_n z^n .
\eeq 
It follows from the details of calculation given in the Appendix 
\beq
\vert I_1 \rangle 
&=&\exp \Biggl\{ 2\ln 2 \left(\frac{p^2}{2} -1 \right) 
\Biggr\}  \nn\\
&&
 \exp  \Biggl\{
\Bigl( \sum_{n, m \ge 1} \frac{1}{2} \bar N_{nm}\,\frac{\a^{\dag}_n}{2} \cdot 
\frac{\a^{\dag}_m}{2} 
+ \sum_{n \ge 1}\bar N_{n0} \frac{\a^{\dag}_n}{2} \cdot \frac{p}{2} \Bigr) 
\Biggr\} \nn\\
&&  
\exp  \Biggl\{
\Bigl( \sum_{n, m \ge 1} \frac{1}{2} \bar N_{nm}\,\frac{\tilde\a^{\dag}_n}{2} \cdot \frac{\tilde\a^{\dag}_m}{2} 
+ \sum_{n \ge 1}\bar N_{n0} \frac{\tilde\a^{\dag}_n}{2} \cdot \frac{p}{2} \Bigr) \Biggr\}\vert 0 \rangle 
\eeq  
where
\beq
\bar N_{00} &=& \ln 2 , ~~ \bar N_{10} = 0, ~~  \bar N_{20} =1, ~~ \bar N_{30} =0, ~~ \bar N_{40} = \half, ~~  \bar N_{50} =0, ~~  \bar N_{60} = - \frac{5}{3}, \cdots, \nn\\
\bar N_{11} &=& 1 , ~~ \bar N_{12} = 0,~~\bar N_{21} = 0 , ~~ \bar N_{22} = 1,~~\bar N_{13} = 0 , ~~ \bar N_{23} = 0, ~~ \bar N_{33} = - \frac{5}{3}, \cdots . 
\eeq 
Here $\bar N^{rs}_{nm}$ is the corresponding Neumann function of open string.  (If we obtained the Neumann function by directly converting the overlapping delta functional, we would have slightly different results. It is because that the string does not propagate freely right after overlapping, precisely speaking. It propagates freely only at the asymptotic region.)
As reported, 
Kawai-Lewellen-Tye (KLT)  \cite{Kawai1986}  factorization occurs even at the level of the one-string identity. 

\subsection{Construction of Two-Closed-String Vertex} 

\begin{figure}[htbp]
\begin {center}
\epsfxsize=0.5\hsize
%
% Specify the picture file name to be included.
\epsfbox{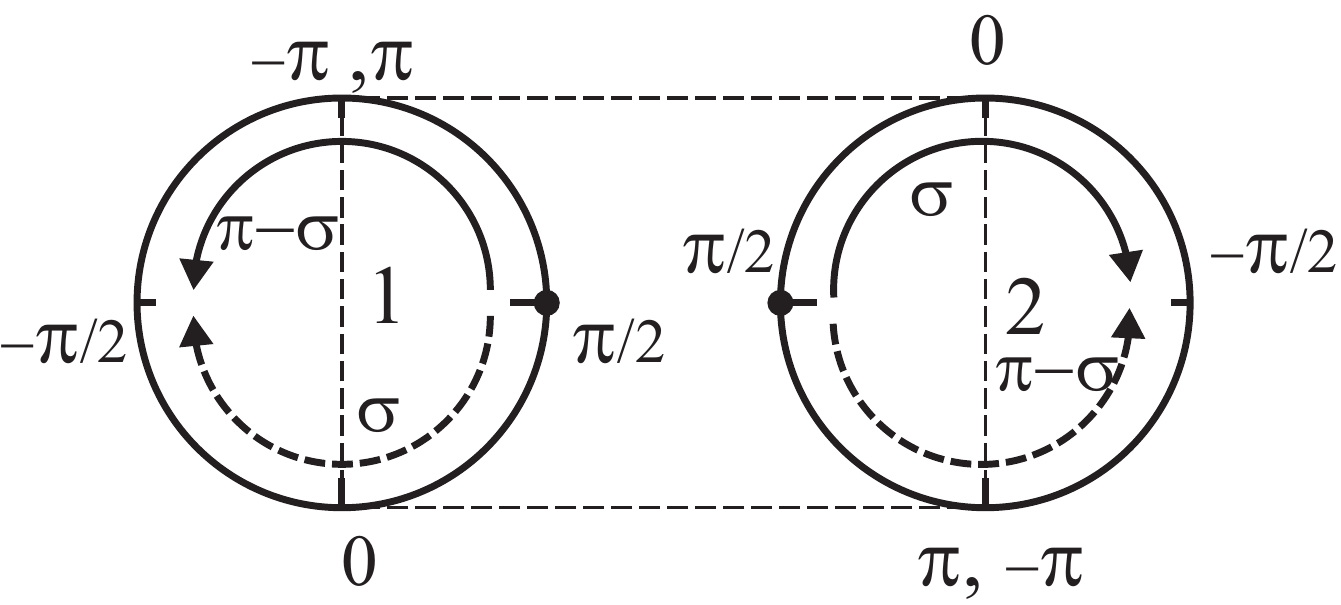}
\end {center}
\caption {\label{double} A Pictorial representation of the two-string identity.}
\end{figure}

Two-string overlapping, which defines the two-string vertex, may be written as 
\beq
\langle X^{(1)}, X^{(2)} \vert I \rangle &=& \prod_{\frac{\pi}{2} \le \s \le \pi}
\prod_{-\pi \le \s \le - \frac{\pi}{2}} \d \left(X^{(1)}(\s) - X^{(2)}(\pi - \s ) \right) \nn\\
&& 
\prod_{\frac{\pi}{2} \le \s \le \pi}
\prod_{-\pi \le \s \le - \frac{\pi}{2}} \d \left(X^{(2)}(\s) - X^{(1)}(\pi - \s ) \right) .
\eeq
Fig. \ref{double} depicts the two-string identity (overlapping) . 
Mapping from the world-sheet coordinates, $\zeta_r = \xi_r + i \eta_r$, $r= 1, 2$ onto the unit
disk is given as follows: 
\begin{subequations}
\beq
\omega_1 &=&  -i \left(\frac{1+ i e^{\zeta_1}}{1 - i e^{\zeta_1}}\right) = -i\left(\frac{1 - e^{2{\xi_1}} + 2i e^{\xi_1} \cos\eta_1}{ 1 + e^{2\xi_1} + 2 e^{\xi_1} \sin^2\eta_1}\right) , \\
\omega_2 &=& i \left(\frac{1+ i e^{\zeta_2}}{1 - i e^{\zeta_2}}\right)= i \left(\frac{1 - e^{2\xi_2} + 2i e^{\xi_2} \cos\eta_2}{ 1 + e^{2\xi_2} + 2 e^{\xi_2}\sin^2\eta_2}\right).
\eeq 
\end{subequations}
where as 
$- \pi \le \eta_1 \le \pi, ~- \pi \le \eta_2 \le \pi $ and 
$ 0 \le \arg \o_1 \le \pi, ~ \pi \le \arg \o_2 \le 2\pi$ .

%%%%%%%%%%%%%%%%%%%%%%%%%%%%%%%%%%%%%%%%%%%%%%%%%
%%%%%%%%%%%%%%%%%%%%%%%%%%%%%%%%%%%%%%%%%%%%%%%%%

\begin{figure}[htbp]
 \begin {center}
 \epsfxsize=0.4\hsize
 %
 % Specify the picture file name to be included.
 \epsfbox{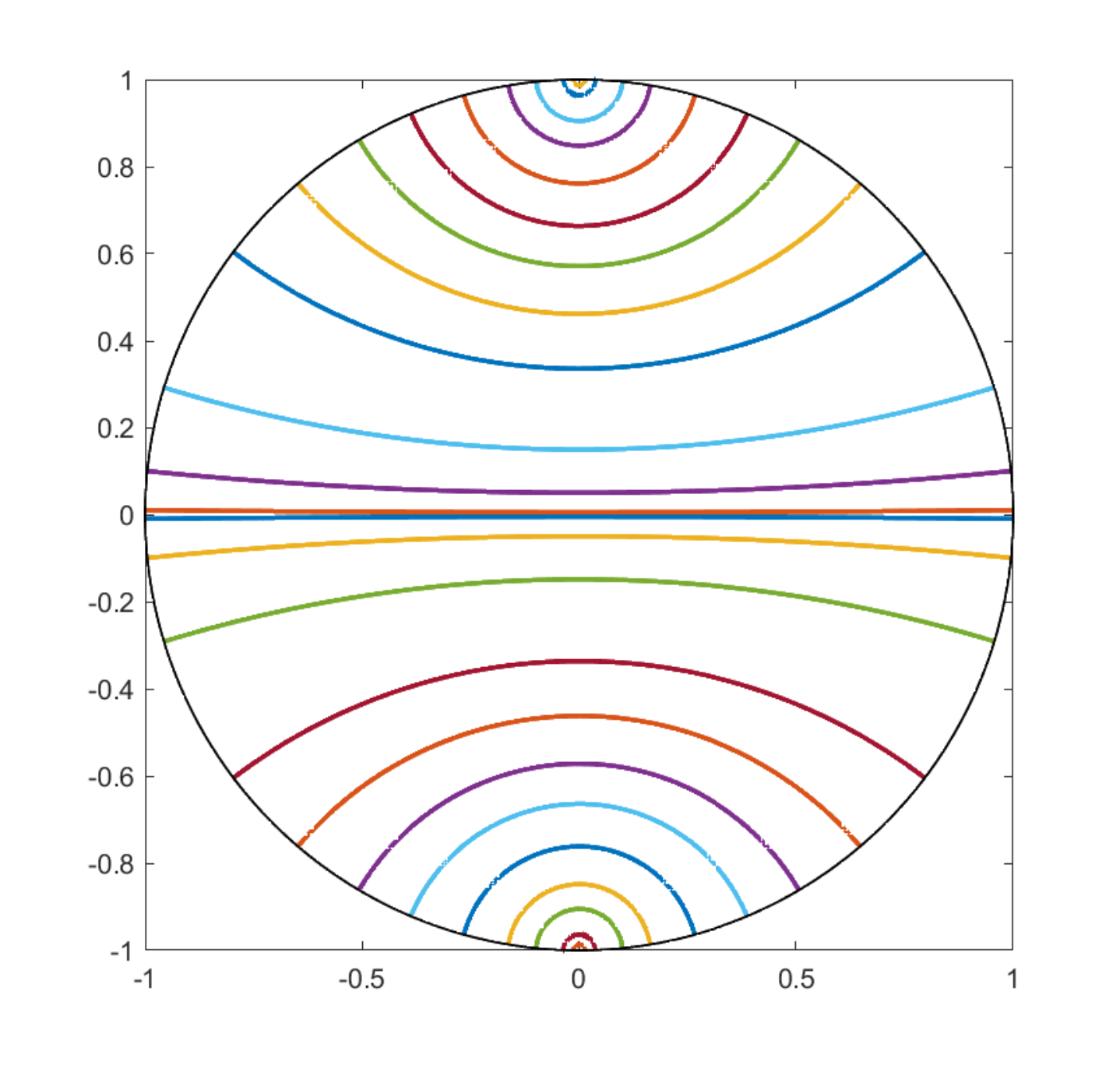}
 \end {center}
 \caption {\label{twostring0pi} Two-string-identity mapped on $\o$-plane (Open string).}
 \end{figure}

\begin{figure}[htbp]
 \begin {center}
 \epsfxsize=0.4\hsize
 %
 % Specify the picture file name to be included.
 \epsfbox{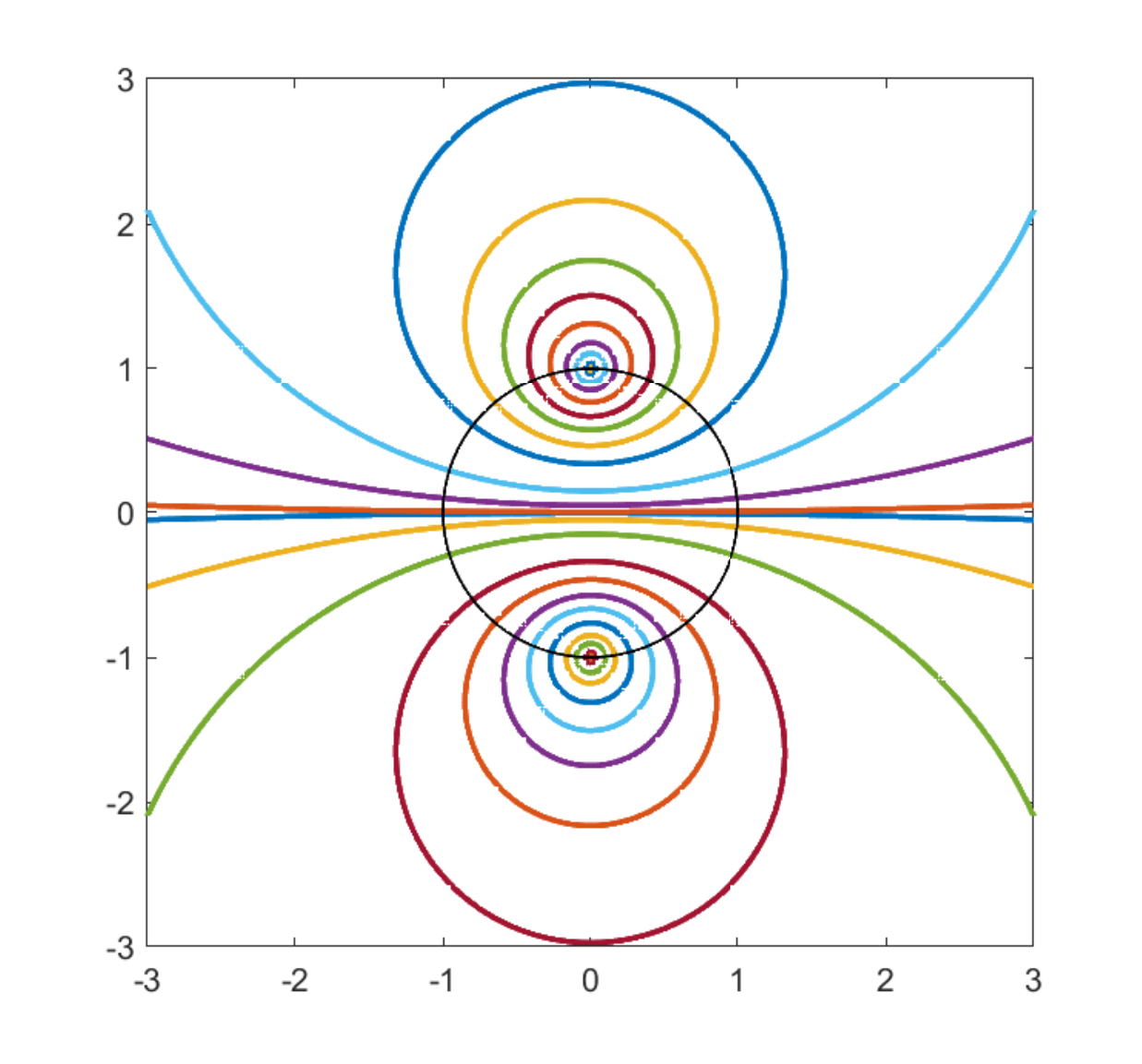}
 \end {center}
 \caption {\label{two-string-pipi} Two-string-identity mapped on $\o$-plane (Closed string).}
 \end{figure}

 Two external strings (at asymptotic regions) are located at $-i$ and $i$ on the unit disk respectively.
 If the ranges of local coordinates $\eta_r$, $r=1,2$ are confined to $[0,\pi]$, the images of string world trajectories are inside of the unit disk. (See Fig. \ref{twostring0pi}.) It corresponds to open string world-sheet. Since 
 \beq
 |\o_r| = \frac{ 1 + e^{2\xi_r} -2 e^{\xi_r} \sin \eta_r}{1 + e^{2\xi_r} +2 e^{\xi_r} \sin \eta_r},
 \eeq 
 and
 \beq 
 |\o_r| \le 1~~~~\text{for} ~~  0 \le \eta \le \pi ~~\text{and} ~~~ |\o_r| \ge 1 ~~~~\text{for}~~ -\pi \le \eta \le 0,
 \eeq 
 if we extend the ranges of $\eta_r$, $r=1,2$ to $[-\pi, \pi]$, the extended parts stretch out the 
 unit disk and form closed curves; therefore, describing the world trajectories of closed strings. 
This is shown in Fig.  \ref{two-string-pipi} ,which depicts $\o$-complex plane.

 %%%%%%%%%%%%%%%%%%%%%%%%%%%%%%%%%%%%%%%%%%%%%%%%%
 %%%%%%%%%%%%%%%%%%%%%%%%%%%%%%%%%%%%%%%%%%%%%%%%%
 
  %%%%%%%%%%%%%%%%%%%%%%%%%%%%%%%%%%%%%%%%%%%%%%%%%
 %%%%%%%%%%%%%%%%%%%%%%%%%%%%%%%%%%%%%%%%%%%%%%%%%

\begin{figure}[htbp]
\begin {center}
\epsfxsize=0.4\hsize
%
% Specify the picture file name to be included.
\epsfbox{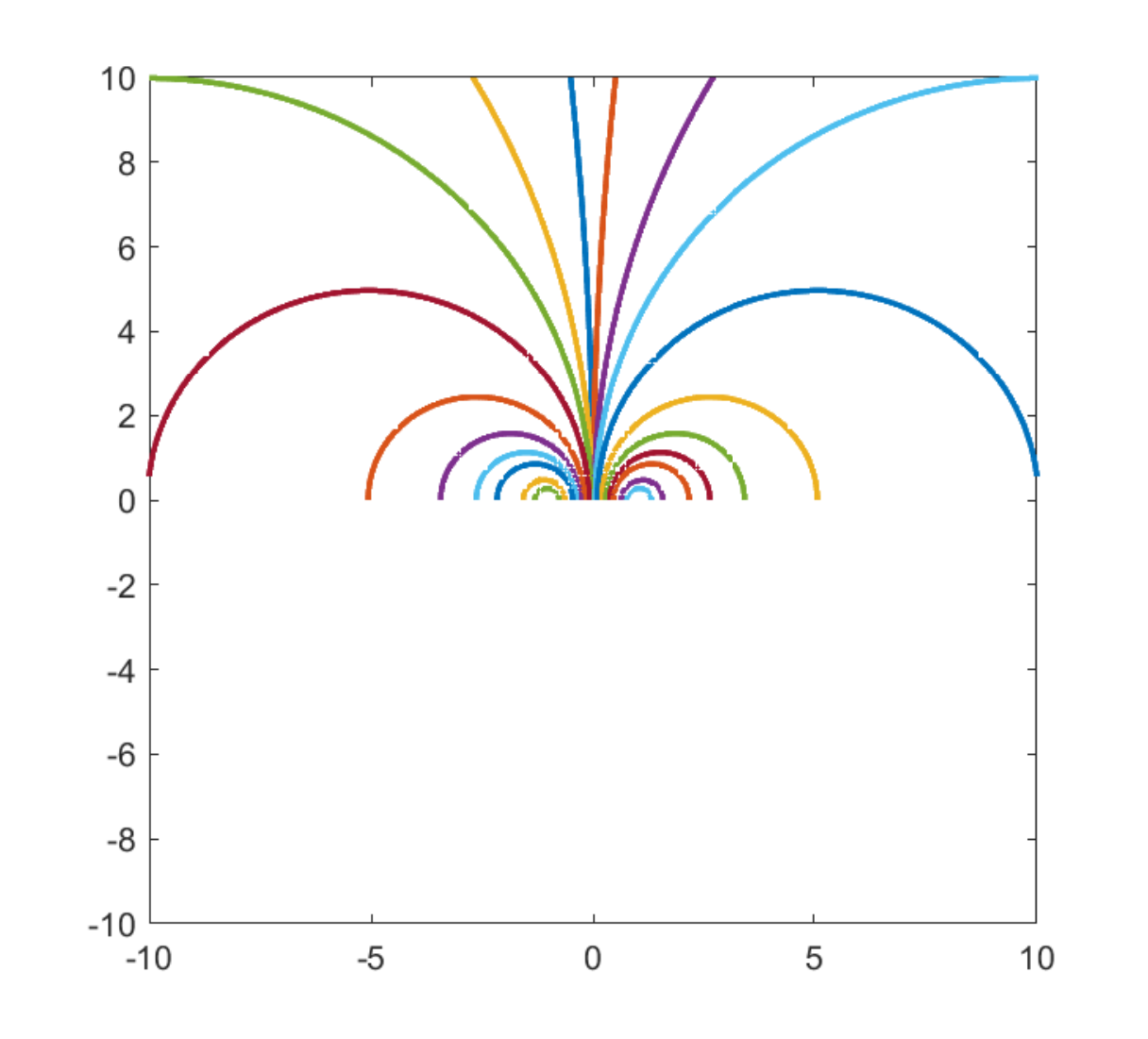}
\end {center}
\caption {\label{twoopenzplane} Two-string-identity mapped on $z$-plane (Open string).}
\end{figure}

\begin{figure}[htbp]
\begin {center}
\epsfxsize=0.4\hsize
%
% Specify the picture file name to be included.
\epsfbox{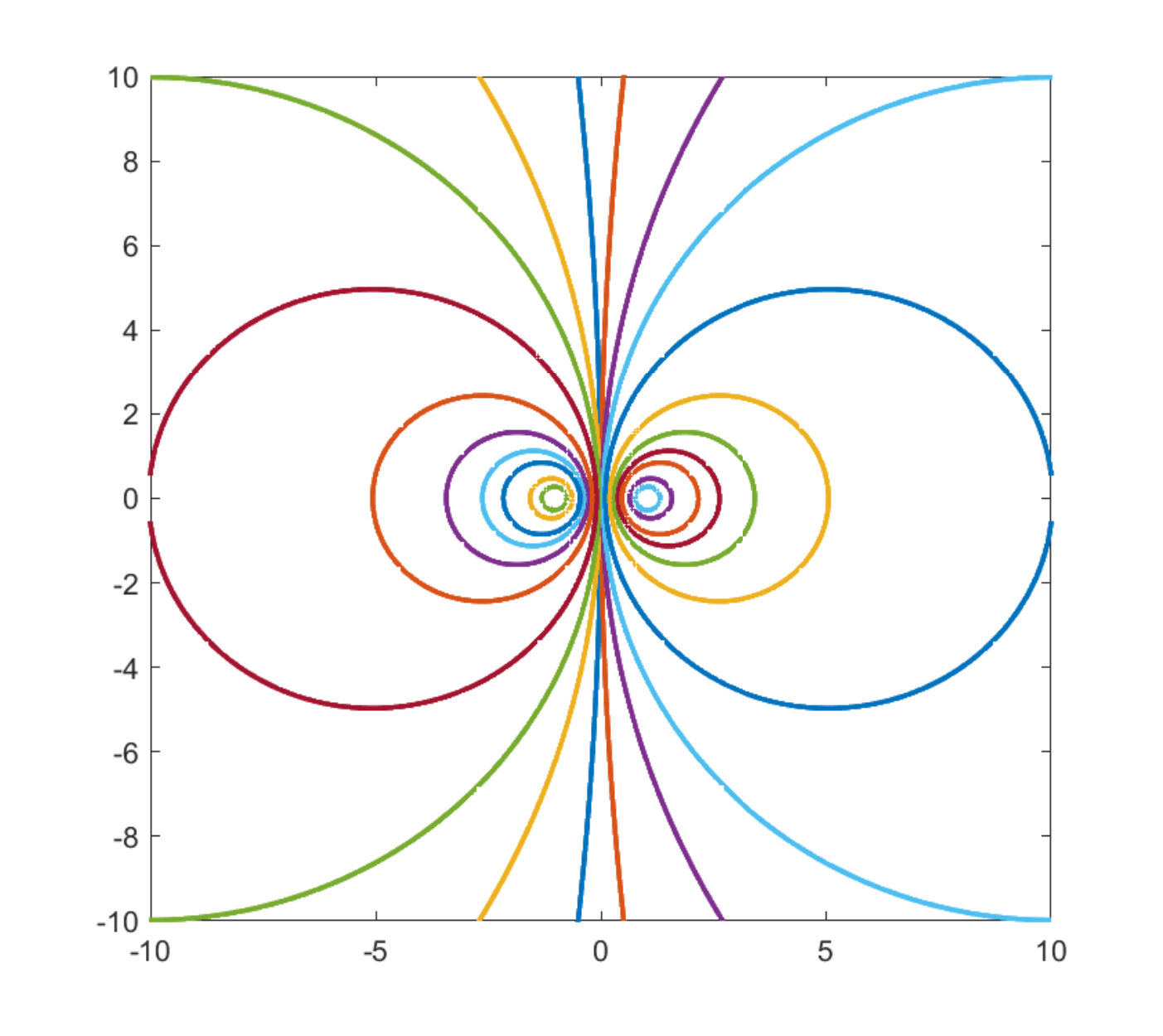}
\end {center}
\caption {\label{twoclosedzplane} Two-string-identity mapped on $z$-plane (Closed string).}
\end{figure}

Mapping onto the complex plane may be carried as follows
\beq
z_1 &=& -i \frac{\o_1 -1}{\o_1 + 1} =  -i \frac{ -i \left(\frac{1+ i e^{\zeta_1}}{1 - i e^{\zeta_1}}\right)  -1}{ -i \left(\frac{1+ i e^{\zeta_1}}{1 - i e^{\zeta_1}}\right)  + 1}, \nn\\
z_2 &=& -i \frac{\o_2 -1}{\o_2 + 1} = -i \frac{i \left(\frac{1+ i e^{\zeta_2}}{1 - i e^{\zeta_2}}\right) -1}{i \left(\frac{1+ i e^{\zeta_2}}{1 - i e^{\zeta_2}}\right) + 1} 
\eeq 
with $\o_r = \left(\frac{1 +iz_r}{1-iz_r} \right)$. 
This maps the external strings $Z_1 = -1 $ and $Z_2 =1$ on the real line of $z$-complex plane
\beq
e^{-\z_1} &=& - i \left(\frac{1+ i\o_1}{1-i \o_1} \right) =  - i \left(\frac{1+ i     \left(\frac{1 +iz_1}{1-iz_1} \right) }{1-i  \left(\frac{1 +iz_1}{1-iz_1} \right)} \right)\nn\\
e^{-\z_2} &=& - i \left(\frac{1- i\o_2}{1+i \o_2} \right) =   - i \left(\frac{1 -i     \left(\frac{1 +iz_2}{1-iz_2} \right) }{1+ i  \left(\frac{1 +iz_2}{1-iz_2} \right)} \right)
\eeq
As in the case of one-string-vertex operator, if the ranges of $\eta_r$ $r=1, 2$ are limited to $[0,\pi]$, the images of the 
string world-sheet covers only the upper half complex $z$-plane  as shown in Fig. \ref{twoopenzplane}. However, with the extended ranges of 
$\eta_r$ $r=1, 2$, the world-sheet covers entire complex $z$-plane. (See Fig. \ref{twoclosedzplane}) 

%%%%%%%%%%%%%%%%%%%%%%%%%%%%%%%%%%%%%%%%%%%%%%%%%
%%%%%%%%%%%%%%%%%%%%%%%%%%%%%%%%%%%%%%%%%%%%%%%%%

%%%%%%%%%%%%%%%%%%%%%%%%%%%%%%%%%%%%%%%%%%%%%%%%%
%%%%%%%%%%%%%%%%%%%%%%%%%%%%%%%%%%%%%%%%%%%%%%%%%

The Fock space representation of the two-closed-string (identity) vertex follows from the general expression of the vertex operator given in Appendix, Eq. (\ref{VNexp} ) 
\beq
\vert I[2] \rangle &=&
\exp \Biggl\{ 2\sum_{r=1}^2 \ln 2 \left( \frac{\left(p^{(r)}\right)^2}{2} -1 \right) 
\Biggr\}~ 2^{2 p^{(1)} \cdot p^{(2)}}  \nn\\
&&
\exp  \Biggl\{
\sum_{r,s} \Bigl( \sum_{n, m \ge 1} \frac{1}{2} \bar N^{rs}_{nm}\,\frac{\a^{(r)\dag}_n}{2} \cdot 
\frac{\a^{(r)\dag}_m}{2} 
+ \sum_{n \ge 1}\bar N^{rs}_{n0} \frac{\a^{(r)\dag}_n}{2} \cdot \frac{p^{(s)}}{2} \Bigr) 
\Biggr\} \nn\\
&& \exp  \Biggl\{
\sum_{r,s} \Bigl( \sum_{n, m \ge 1} \frac{1}{2} \bar N^{rs}_{nm}\,\frac{\tilde\a^{(r)\dag}_n}{2} \cdot \frac{\tilde\a^{(r)\dag}_m}{2} 
+ \sum_{n \ge 1}\bar N^{rs}_{n0} \frac{\tilde\a^{(r)\dag}_n}{2} \cdot \frac{p^{(s)}}{2} \Bigr) \Biggr\}
\vert 0 \rangle . 
\eeq 
Here the Neumann functions are explicitly evaluated as 
\beq
\bar N^{12}_{00} &=&   \ln \vert Z_1 - Z_2 \vert = \ln 2, ~~\bar N^{11}_{00} = \bar N^{22}_{00} = \ln 2\nn\\
N^{11}_{10} &=& -1, ~~N^{22}_{10} =1, ~~ N^{12}_{10}= -1, ~~ N^{21}_{10} = 1,\nn\\
N^{11}_{20} &=& \half, ~~N^{22}_{20} =\half, ~~ N^{12}_{20}= 1, ~~ N^{21}_{20} = 1,\nn\\
N^{11}_{30} &=& - \frac{1}{3}, ~~N^{22}_{30} =\frac{1}{3}, ~~ N^{12}_{30}= -1, ~~ N^{21}_{30} = 1.\nn\\
\eeq

\section{ Construction of Cubic Closed String Vertex} 

\begin{figure}[htbp]
\begin {center}
\epsfxsize=0.9\hsize
%
% Specify the picture file name to be included.
 \epsfbox{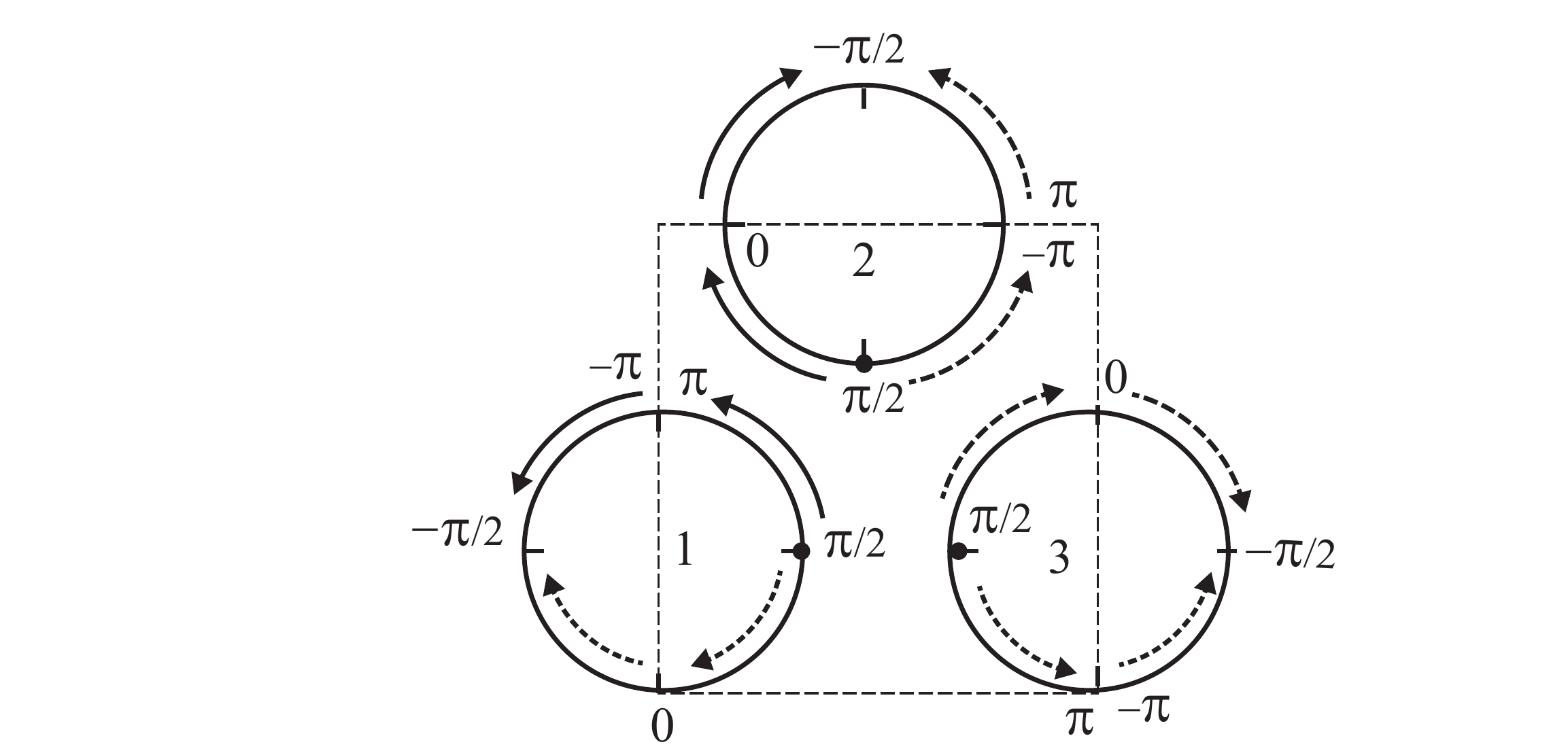}
\end {center}
\caption {\label{starclosed} Star product of three closed strings.}
\end{figure}

\begin{figure}[htbp]
%\begin {center}
%\input epsf.tex
\epsfxsize=0.5\hsize
%
% Specify the picture file name to be included.
\epsfbox{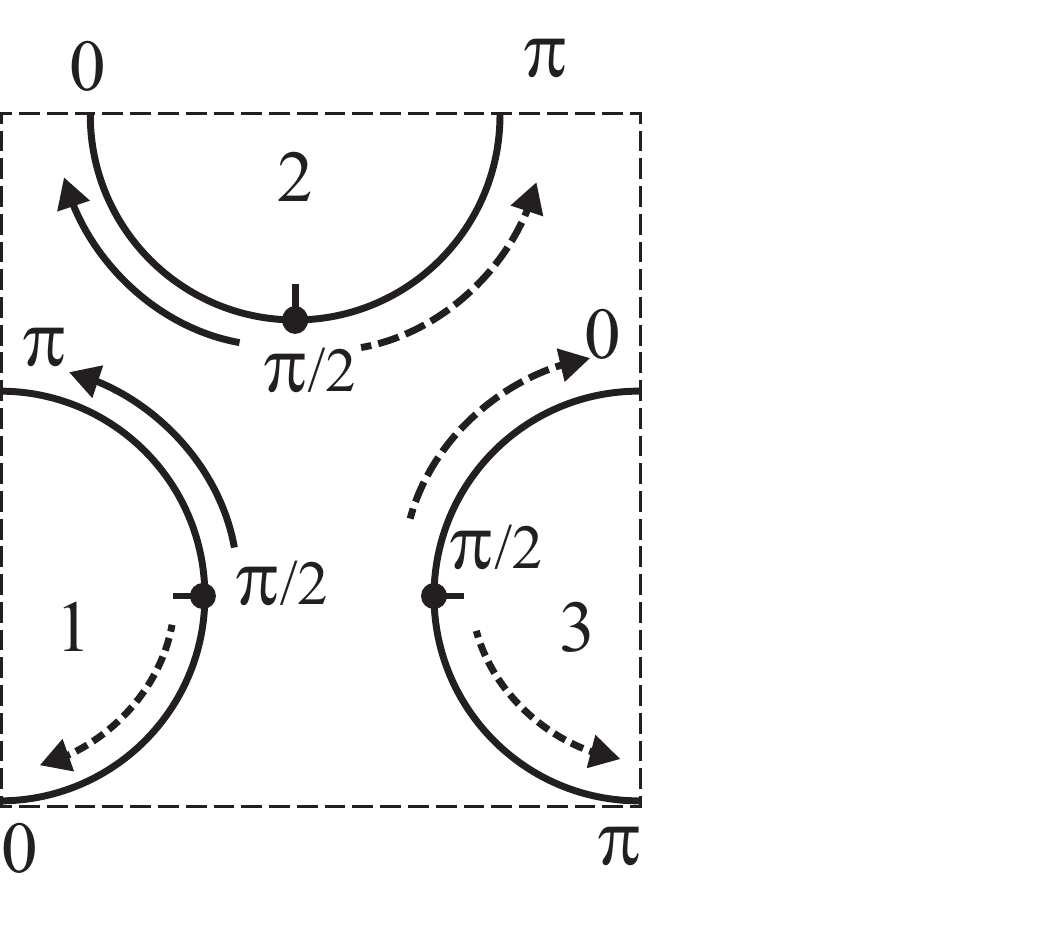}
%\end {center}
\caption {\label{staropen2} Star product of three open strings.}
\end{figure}

 The overlapping condition for three-closed-string interaction is expressed as 
\beq
X^{(r)}(\s) = X^{(r+1)}(\pi -\s),~~~\text{for}~~ \frac{\pi}{2} \le \s \le \pi, ~\text{and}~
-\pi \le \s \le - \frac{\pi}{2}, ~~~r = 1, 2, 3 .
\eeq 
where we identify $X^{(r+3)} = X^{(r)}$.  In accordance with it, the star product between two string field operators may be defined as 
\beq
\left(\Psi_1*\Psi_2\right)[X(\s)]_{\rm closed} &=&  \int\prod_{-\pi \le \s \le - \frac{\pi}{2}} \prod_{\frac{\pi}{2} \le \s \le \pi} DX^{(1)}(\s) \prod_{0 \le \s \le \frac{\pi}{2}}
\prod_{-\frac{\pi}{2} \le \s \le 0} DX^{(2)}(\s)  \nn\\
&& 
\prod_{ -\pi \le \s \le - \frac{\pi}{2}} \prod_{\frac{\pi}{2} \le \s \le \pi} \d \left[X^{(1)}(\s) - X^{(2)}(\pi -\s) \right] \Psi_1[X^{(1)}(\s)] \Psi_2[X^{(2)}(\s)], \ 
\eeq 
It would be instructive to compare it with the previous star product of open string defined by Witten 
\beq
\left(\Psi_1*\Psi_2\right)[X(\s)]_{\rm open} &=&  \int\prod_{\frac{\pi}{2} \le \s \le \pi} DX^{(1)}(\s) \prod_{0 \le \s \le \frac{\pi}{2}} DX^{(2)}(\s)  \nn\\
&& 
\prod_{\frac{\pi}{2} \le \s \le \pi} \d \left[X^{(1)}(\s) - X^{(2)}(\pi -\s) \right] \Psi_1[X^{(1)}(\s)] \Psi_2[X^{(2)}(\s)], \ 
\eeq
The star product of three closed strings were illustrated in Fig. \ref{starclosed}. It may be instructive to compare this star product of closed string with other previously reported star product of Witten open string presented in Fig. \ref{staropen2}. We recognize that the region inside the dotted box of Fig. \ref{starclosed} of closed string corresponds to the
star product of the open string.  

%%%%%%%%%%%%%%%%%%%%%%%%%%%%%%%%%%%%%%%%%%%%%%%%%
%%%%%%%%%%%%%%%%%%%%%%%%%%%%%%%%%%%%%%%%%%%%%%%%%

%%%%%%%%%%%%%%%%%%%%%%%%%%%%%%%%%%%%%%%%%%%%%%%%%
%%%%%%%%%%%%%%%%%%%%%%%%%%%%%%%%%%%%%%%%%%%%%%%%%

Now, being equipped with the proper star product for the closed strings, we may write the BRST gauge invariant 
action for closed string:
\beq
S_{\rm Closed} = \int \tr \left( \Psi * \left(Q + \tilde Q\right)\Psi + \frac{2g}{3} \Psi * \Psi * \Psi \right),
\eeq 
which is invariant under the (extended) BRST gauge transformation 
\beq
\d \Psi = Q* \e + \Psi * \e - \e * \Psi  + \tilde Q* \tilde\e + \Psi * \tilde\e - \tilde \e * \Psi  .
\eeq

Mapping from the world-sheet coordinates, $\zeta_r = \xi_r + i \eta_r$, $r= 1, 2, 3$ to the 
disk is given as follows: 
\begin{subequations}
\beq
\omega_1 &=& e^{\frac{2\pi i}{3}} \left(\frac{1+ i e^{\zeta_1}}{1 - i e^{\zeta_1}}\right)^{\frac{2}{3}}, \\
\omega_2 &=& \left(\frac{1+ i e^{\zeta_2}}{1 - i e^{\zeta_2}}\right)^{\frac{2}{3}},\\
\omega_3 &=& e^{-\frac{2\pi i}{3}} \left(\frac{1+ i e^{\zeta_3}}{1 - i e^{\zeta_3}}\right)^{\frac{2}{3}}
\eeq 
\end{subequations}
A pair of two string patches meet along the line $\xi_r$, $r=1,2,3$ 
and the three string patches meet at $(\xi_r, \eta_r) = (0, \frac{\pi}{2}), ~r=1,2,3$. 
If we choose the domains of $\eta_r$, $r=1,2,3$, as $[0,\pi]$, $\o$, $r=1,2,3$ describe the world-sheet of three open strings as given in Fig. \ref{staropenw}. External string are at the region where $\xi_r \rightarrow -\infty$: $e^{\frac{2\pi i}{3}}$, $1$, $ e^{-\frac{2\pi i}{3}} $ on $\o$-complex plane. The images of the string world-sheet remain inside of the unit disk. 
\beq 
|\o_r| \le 1~~~~\text{for} ~~  0 \le \eta_r \le \pi, ~~r = 1, 2, 3. 
\eeq 
However, if the domains of $\eta_r$, $r=1,2,3$ are extended to $[-\pi, \pi]$, their images stretch out the unit disk as earlier shown  in Fig. \ref{starclosedw} :
\beq
|\o_r| \ge 1 ~~~~\text{for}~~ -\pi \le \eta_r \le 0, ~~r = 1, 2, 3.
\eeq 

\begin{figure}[htbp]
%\begin {center}
%\input epsf.tex
\epsfxsize=0.5\hsize
%
% Specify the picture file name to be included.
\epsfbox{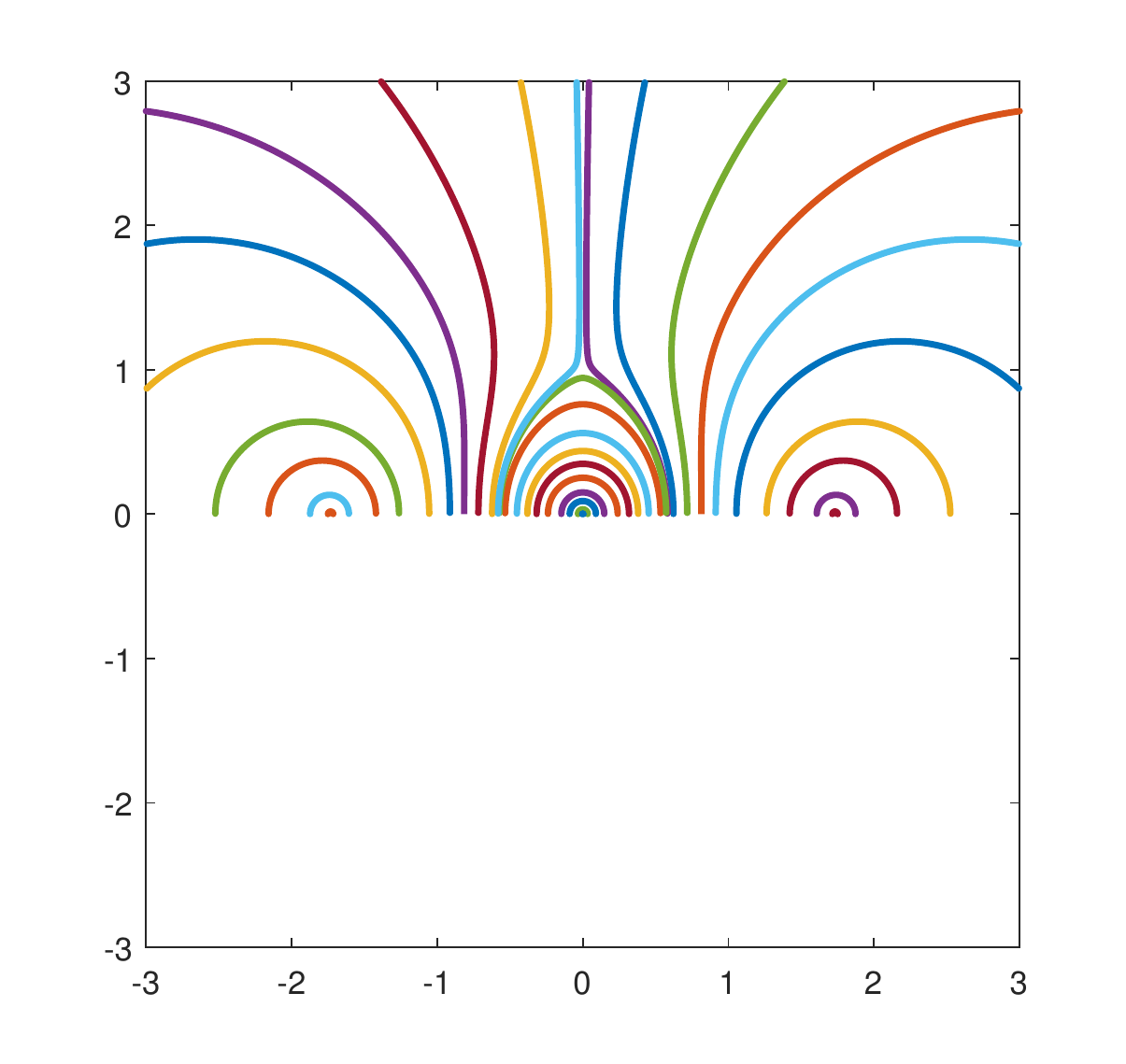}
%\end {center}
\caption {\label{3stringz-planeopen} Cubic open string world-sheet on complex $z$-plane.}
\end{figure}

\begin{figure}[htbp]
%\begin {center}
%\input epsf.tex
\epsfxsize=0.5\hsize
%
% Specify the picture file name to be included.
\epsfbox{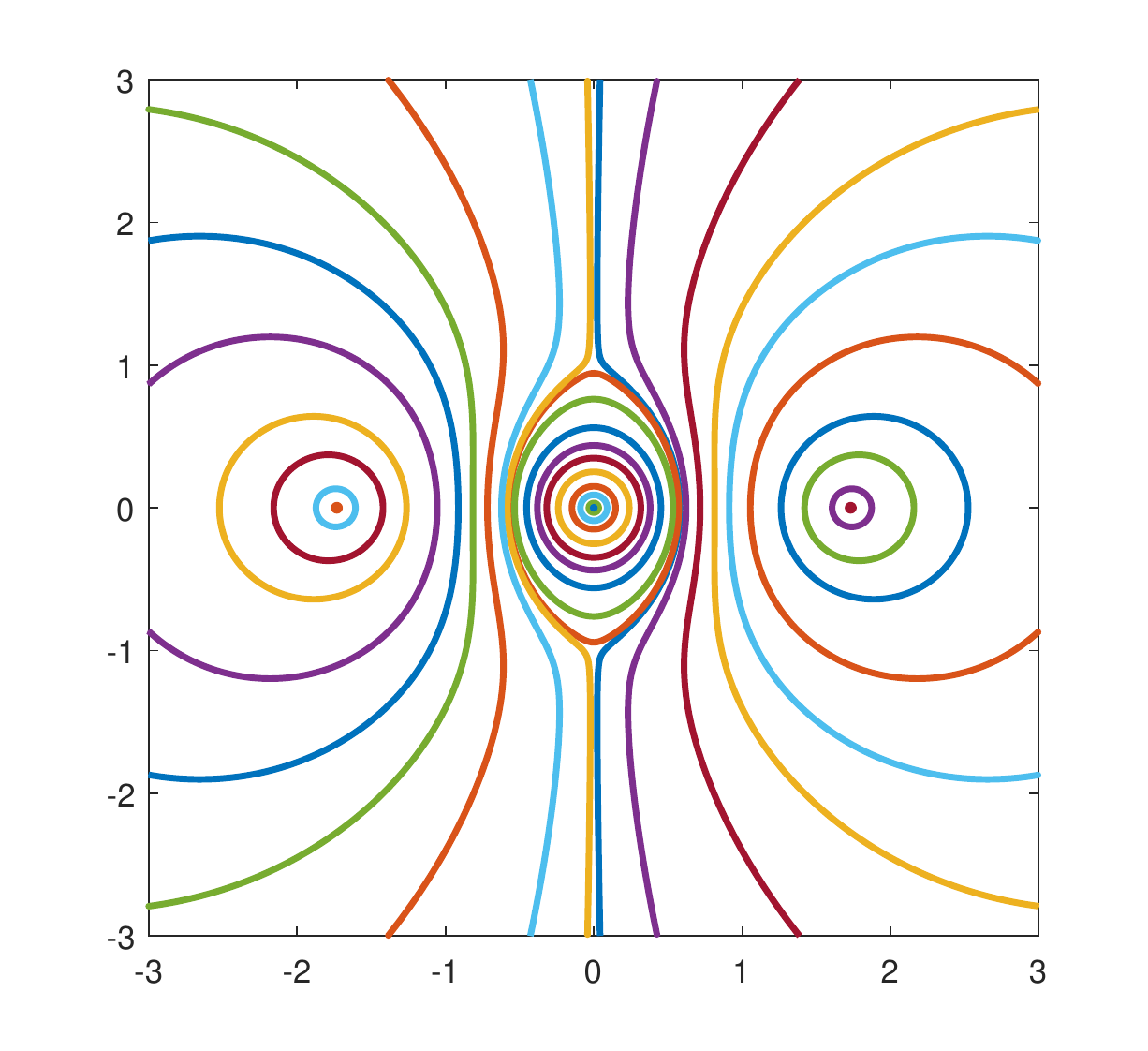}
%\end {center}
\caption {\label{3stringz-planeclosed} Cubic closed string world-sheet on complex $z$-plane.}
\end{figure}

The conformal transformation we employed before to map the 
complex $\o$-plane to the complex $z$-plane: 
\beq
z_r = -i \,\frac{\omega_r -1}{\omega_r +1}, ~~ r =1, ~2,~ 3.
\eeq 
The external strings are now located on the real line 
\beq
Z_1 = \sqrt{3}, ~~ Z_2 = 0, ~~~ Z_3 = - \sqrt{3}.
\eeq 
Each local coordinate patch is mapped onto the upper half plane for 
$ 0 \le \eta_r \le \pi, ~~r = 1, 2, 3$ (open string, see Fig. \ref{3stringz-planeopen} and onto the lower half plane for 
$-\pi \le \eta_r \le 0, ~~r = 1, 2, 3$. Thus, string world-sheet covers the entire complex $z$-plane for closed string (Fig. \ref{3stringz-planeclosed}) 

%%%%%%%%%%%%%%%%%%%%%%%%%%%%%%%%%%%%%%%%%%%%%%%%%
%%%%%%%%%%%%%%%%%%%%%%%%%%%%%%%%%%%%%%%%%%%%%%%%%

%%%%%%%%%%%%%%%%%%%%%%%%%%%%%%%%%%%%%%%%%%%%%%%%%
%%%%%%%%%%%%%%%%%%%%%%%%%%%%%%%%%%%%%%%%%%%%%%%%%

The relations between the local coordinates and the $z$-complex coordinates may be manifested through expansions of  $e^{-\z_r}$, $r=1,2,3$ near $z_r = Z_r$ (at the asymptotic region),
\beq
e^{-\z_r} &=& \frac{a_r}{(z_r-Z_r)} + \sum_{n=0} c^{(r)}_n (z_r-Z_r)^n , \label{expand46}\\
a_1 &=& \frac{8}{3}, ~~ a_2 = \frac{2}{3}, ~~~ a_3 = \frac{8}{3}, \nn\\
c^{(1)}_0 &=&\frac{2\sqrt{3}}{3}, ~~ c^{(1)}_1 = - \frac{5}{72} , ~~c^{(1)}_2 = \frac{5\sqrt{3}}{288} \nn\\
c^{(2)}_0 &=& 0, ~~ c^{(2)}_1 = - \frac{5}{18}, ~~ c^{(2)}_2 = 0 ,\nn\\
c^{(3)}_0 &=& -\frac{2\sqrt{3}}{3}, ~~ c^{(3)}_1 = - \frac{5}{72} , ~~c^{(3)}_2 = -\frac{5\sqrt{3}}{288} . \nn %\label{expand47}
\eeq
These explicit expressions of expansions are useful when we calculating the Neumann functions. The general 
expression of the $N$-vertex operator for the closed string is presented at the end of Appendix Eq. (\ref{VNexp}),
where the KLT relations are manifestly built in.

\section{Three-Graviton Scattering Amplitude}

It is important to check if the constructed closed string field theory is compatible with the 
Einstein gravity. The first step will be to reproduce the three-graviton scattering amplitude. 
For the case of the closed string field theory in the proper-time gauge \cite{TLeeEPJ2018}, 
it has been confirmed that the constructed covariant closed string field theory reproduces the three-graviton scattering amplitude correctly and the KLT relations are manifestly encoded in the scattering amplitude. 
The general expression of $N$-vertex operator has been given in Appendix Eq. (\ref{VNexp}) in terms of 
the Neumann functions. To evaluate the three-graviton scattering amplitude 
we need the Neumann functions of $\bar N^{rs}_{nm} = \bar N^{sr}_{mn}$, $r, s = 1, 2, 3$, $n, m \le 1$.
It involved some algebra to calculate them explicitly. We evaluated Neumann functions of $\bar N^{rs}_{00}$, using Eq. (\ref{a33a}) and Eq. (\ref{a33b}) given in Appendix,
\beq \label{N00}
\bar N^{11}_{00} &=& \ln \frac{8}{3}, ~~\bar N^{22}_{00} = \ln \frac{2}{3}, ~~ \bar N^{33}_{00} = \ln \frac{8}{3} ,\nn\\
\bar N^{12}_{00} &=& \frac{1}{2}\ln 3, ~~\bar N^{23}_{00} = \frac{1}{2} \ln 3, ~~ \bar N^{31}_{00} =  \ln 2 \sqrt{3} . 
\eeq 
The Neumann functions of $\bar N^{rs}_{10}$ follows from Eq. (\ref{expand46}) and Eq. (\ref{a33c}):
\beq
N^{rr}_{10} = c^{(r)}_0   ,~~~\bar N^{rs}_{10} = \frac{a_r}{(Z_r-Z_s)}, ~~\text{for}~~r\not=s . 
\eeq 
If we spell them out: 
\beq \label{N10}
\bar N^{11}_{10} &=&\frac{2\sqrt{3}}{3}, ~~\bar N^{22}_{10}= 0 , ~~\bar N^{33}_{10}= - \frac{2\sqrt{3}}{3}, \nn\\ 
\bar N^{12}_{10} &=& \frac{8}{3\sqrt{3}}, ~~ \bar N^{13}_{10} =  \frac{4}{3\sqrt{3}}, ~~
\bar N^{21}_{10} = - \frac{2}{3 \sqrt{3}}, \\
\bar N^{23}_{10} &=& \frac{2}{3\sqrt{3}}, ~~ \bar N^{31}_{10} = -  \frac{4}{3\sqrt{3}}, ~~
\bar N^{32}_{10} = - \frac{8}{3\sqrt{3}} .  \nn
\eeq

The Neumann functions of $\bar N^{rs}_{11}$ may be obtained from the general formula 
Eq. (\ref{a33d}) for $\bar N^{rs}_{nm}$:
\beq \label{Nnm}
\bar N^{rs}_{mn} =\frac{1}{nm} \oint_{Z_r} \frac{dz_r}{2\pi i} \oint_{Z_s} \frac{d z^\prime_s}{2\pi i} 
\frac{1}{(z_r-z^\prime_s)^2} e^{-n\zeta_r(z_r) - m \zeta^\prime_s(z^\prime_s)}, ~~~ n, m \ge 1
\eeq 
For the case with $n=1$ and $m=1$:
\beq
\bar N^{rs}_{11} &=& \oint_{Z_r} \frac{dz_r}{2\pi i} \oint_{Z_s} \frac{d z^\prime_s}{2\pi i} 
\frac{1}{(z_r-z^\prime_s)^2} \left( \frac{a_r}{z_r-Z_r}\right)\left( \frac{a_s}{z^\prime_s-Z_s}\right). 
\eeq 
When $r\not= s$, 
\beq
\bar N^{rs}_{11} &=& \frac{a_r a_s}{(Z_r -Z_s)^2} = \frac{2^4}{3^3} , ~~~{\rm for}~~ r, s = 1,2,3.
\eeq 
When $r=s$, we may write 
\beq
\bar N^{rr}_{11} &=& \oint_{Z_r} \frac{dz_r}{2\pi i} \oint_{Z_s} \frac{d z^\prime_r}{2\pi i} 
\frac{1}{(z_r-z^\prime_r)^2} \left( \frac{a_r}{z-Z_r}+ \sum_{m=0} c^{(r)}_m (z_r-Z_r)^m \right)\left( \frac{a_r}{z^\prime_r-Z_r}+ \sum_{m=0} c^{(r)}_m (z_r-Z_r)^m \right) \nn\\
&=& a_r c^{(r)}_1.
\eeq 
Using Eq. (\ref{expand46}) and 
\beq
c^{(1)}_1 &=& - \frac{5}{72}, ~~c^{(2)}_1 = - \frac{5}{18},~~ c^{(3)}_1 = - \frac{5}{72}, 
\eeq 
we find 
\beq
\bar N^{11}_{11} =\bar N^{22}_{11} =\bar N^{33}_{11} = - \frac{5}{27}. 
\eeq

With the Neumann functions for three closed string Eq. (\ref{N00}), Eq. (\ref{N10}), and Eq. (\ref{Nnm}), the three-closed-string vertex operator may be expressed as 
\beq \label{vertexclosed3}
\vert V^{\text{Closed}}_{[3]}[1,2,3] \vert 0 \rangle 
&=&  \exp \Biggl\{ 2\sum_{r=1}^3 \ln \frac{8}{3} \left( \frac{\left(p^{(r)}\right)^2}{2} -1 \right) 
\Biggr\}\prod_{r <s} \vert Z_r - Z_s \vert^{2 p^{(r)} \cdot p^{(s)}}~\nn\\
&&
\exp  \Biggl\{
\sum_{r,s} \Bigl( \sum_{n, m \ge 1} \frac{1}{2} \bar N^{rs}_{nm}\,\frac{\a^{(r)\dag}_n}{2} \cdot 
\frac{\a^{(s)\dag}_m}{2} 
+ \sum_{n \ge 1}\bar N^{rs}_{n0} \frac{\a^{(r)\dag}_n}{2} \cdot \frac{p^{(s)}}{2} \Bigr) 
\Biggr\} \nn\\
&& \exp  \Biggl\{
\sum_{r,s} \Bigl( \sum_{n, m \ge 1} \frac{1}{2} \bar N^{rs}_{nm}\,\frac{\tilde\a^{(r)\dag}_n}{2} \cdot \frac{\tilde\a^{(r)\dag}_m}{2} 
+ \sum_{n \ge 1}\bar N^{rs}_{n0} \frac{\tilde\a^{(r)\dag}_n}{2} \cdot \frac{p^{(s)}}{2} \Bigr) \Biggr\} \vert 0 \rangle
\eeq 
Note that the scattering amplitude of three closed strings can be factorized into those of three open strings except for the zero modes (the KLT relations). The $Z_r$ dependent term may be absorbed into the $SL(2,R)$ invariant measure.  In fact, we can remove the $Z_r$ dependent term, $\prod_{r <s} \vert Z_r - Z_s \vert^{2 p^{(r)} \cdot p^{(s)}} $, for graviton
scattering on shell, since
\beq
2 p^{(r)} \cdot p^{(s)} = (p^{(r)})^2 + 2 p^{(r)} \cdot p^{(s)} + (p^{(s)})^2 = \left( p^{(r)} + p^{(s)} \right)^2
=  (p^{(t)})^2 = 0, 
\eeq 
where $t \not= r~{\rm or} ~ s$. 
The momentum dependent terms may be written as 
\beq
\sum_{r, s} \bar N^{rs}_{10}~ \a^{(r)\dag}_1 \cdot p^{(s)} &=& 
\sum_r\bar N^{rr}_{10}~ \a^{(r)\dag}_1 \cdot p^{(r)} + \sum_{r, s}{}^\prime\bar N^{rs}_{10}~ \a^{(r)\dag}_1 \cdot p^{(s)} \nn\\
&=& \frac{2\sqrt{3}}{3} \a^{(1)\dag}_1 \cdot p^{(1)} - \frac{2\sqrt{3}}{3} \a^{(3)\dag}_1 \cdot p^{(3)} +
\frac{4}{3 \sqrt{3}}\a^{(1)\dag}_1 (p^{(2)}- p^{(1)}) \nn\\
&& + \frac{2}{3\sqrt{3}} \a^{(2)\dag}_1 \cdot  (p^{(3)}- p^{(1)}) + \frac{4}{3\sqrt{3}} \a^{(2)\dag}_1 \cdot  (p^{(3)}- p^{(2)})
\eeq 

Choosing the external string state as 
\beq
\vert \Psi_1, \Psi_2, \Psi_3 \rangle = \prod_{r=1}^3 \left\{h_{\m\n}(p^r) \a^{(r)\m}_{-1} \tilde \a^{(r)\n}_{-1} \right\}\vert 0, \rangle 
\eeq 
we can obtain the three graviton scattering amplitude out of the three closed string scattering amplitude 
\beq
{\cal A}_{[3]} &=& \int \prod_{r=1}^3 dp^{(r)} \d \left(\sum_{r=1}^3 p^{(r)}\right) \frac{2g}{3} \, \langle \Psi^{(1)}, \Psi^{(2)}, \Psi^{(3)} \vert V^{\text{Closed}}_{[3]}[1,2,3] \vert 0 \rangle \nn\\
&=& \left(\frac{2g}{3}\right) \left(\frac{3}{8}\right)^6
 \int \prod_{i=1}^3 dp^{(i)} \d\left(
\sum_{i=1}^3 p^{(i)} \right) \langle 0 \vert \left\{\prod_{i=1}^3 h_{\m\n}(p^{(i)}) a^{(i)\m}_1 \cdot \tilde a^{(i)\n}_1 \right\} \frac{1}{2^5}\left(\sum_{r, s =1}^3 \bar N^{rs}_{11} a^{(r)\dag}_1 \cdot a^{(s)\dag}_1 \right) \nn\\
&& \left(\sum_{t,u} \bar N^{tu}_{10}~ \a^{(t)\dag}_1 \cdot p^{(u)}\right) \frac{1}{2^5}
\left(\sum_{l, m =1}^3 \bar N^{lm}_{11} \tilde a^{(l)\dag}_1 \cdot \tilde a^{(m)\dag}_1 \right) \left(\sum_{p,q} \bar N^{pq}_{10}~ \a^{(p)\dag}_1 \cdot p^{(q)}\right)  \vert 0 \rangle .
\eeq 
where $a_n = \a_n/\sqrt{n}, ~ a^\dag_n = \a_{-n}/\sqrt{n}$ for $n>0$. 
Using on-shell or in the zero-slope limit, 
\beq
\left(p^{(r)}\right)^2 =0, ~~~ r = 1, 2, 3,
\eeq
and the momentum conservation
\beq
p^{(1)} + p^{(2)}+ p^{(3)} =0,
\eeq
we may simplify the three-graviton-scattering amplitude as 
\beq
{\cal A}_{[3]} &=&  \left(\frac{2g}{3}\right) \left(\frac{3}{8}\right)^6 \frac{1}{2^{10}} \frac{2}{3\sqrt{3}}
\int \prod_{i=1}^3 dp^{(i)} \d\left( \sum_{i=1}^3 p^{(1)} \right) \nn\\
&& h_{\m_1\n_1}(p^{(1)})\,h_{\m_2\n_2}(p^{(2)})\,h_{\m_3\n_3}(p^{(3)})\biggl\{\bar N^{12}_{11} \eta^{\m_1\m_2}\left(p^{(1)} - p^{(2)}\right)^{\m_3}
+ \bar N^{13}_{11} \eta^{\m_1\m_3}\left(p^{(3)} - p^{(1)}\right)^{\m_2} \nn\\
&& \bar N^{21}_{11} \eta^{\m_2\m_1}\left(p^{(1)} - p^{(2)}\right)^{\m_3} + \bar N^{23}_{11} \eta^{\m_2\m_3}\left(p^{(2)} - p^{(3)}\right)^{\m_1}
+ \bar N^{31}_{11} \eta^{\m_3\m_1}\left(p^{(3)} - p^{(1)}\right)^{\m_2} \nn\\
&& + \bar N^{32}_{11} \eta^{\m_3\m_2}\left(p^{(2)} - p^{(3)}\right)^{\m_1} \biggr\}          \biggl\{\bar N^{12}_{11} \eta^{\n_1\n_2}\left(p^{(1)} - p^{(2)}\right)^{\n_3}+ \bar N^{13}_{11} \eta^{\n_1\n_3}\left(p^{(3)} - p^{(1)}\right)^{\n_2} \nn\\
&& \bar N^{21}_{11} \eta^{\n_2\n_1}\left(p^{(1)} - p^{(2)}\right)^{\n_3} + \bar N^{23}_{11} \eta^{\n_2\n_3}\left(p^{(2)} - p^{(3)}\right)^{\n_1}
+ \bar N^{31}_{11} \eta^{\n_3\n_1}\left(p^{(3)} - p^{(1)}\right)^{\n_2} \nn\\
&& + \bar N^{32}_{11} \eta^{\n_3\n_2}\left(p^{(2)} - p^{(3)}\right)^{\n_1} \biggr\}
\eeq 
Here we make use of 
\beq
\bar N^{11}_{10} p^{(1)} + N^{12}_{10} p^{(2)} + N^{13}_{10} p^{(3)} &=& \frac{2}{3\sqrt{3}} \left(p^{(2)} - p^{(3)} \right) \nn\\
\bar N^{21}_{10} p^{(1)} + N^{22}_{10} p^{(2)} + N^{23}_{10} p^{(3)} &=& \frac{2}{3\sqrt{3}} \left(p^{(3)} - p^{(1)} \right) \nn\\
\bar N^{31}_{10} p^{(1)} + N^{32}_{10} p^{(2)} + N^{33}_{10} p^{(3)} &=& \frac{2}{3\sqrt{3}} \left(p^{(1)} - p^{(2)} \right) .
\eeq 
Using the momentum conservation and the covariant gauge condition, we further simplify the 
three-graviton-scattering amplitude ${\cal A}_{[3]} $:
\beq \label{threegrav}
{\cal A}_{[3]}  
&=& \kappa
\int \prod_{i=1}^3 dp^{(i)} \d\left( \sum_{i=1}^3 p^{(i)} \right) \nn\\
&& h_{\m_1\n_1}(p^{(1)})\,h_{\m_2\n_2}(p^{(2)})\,h_{\m_3\n_3}(p^{(3)})\biggl\{\eta^{\m_1\m_2}\left(p^{(1)} - p^{(2)}\right)^{\m_3} +\eta^{\m_2\m_3}\left(p^{(2)} - p^{(3)}\right)^{\m_1}
+\eta^{\m_3\m_1}\left(p^{(3)} - p^{(1)}\right)^{\m_2} \biggr\}         \nn\\
&& \biggl\{\eta^{\n_1\n_2}\left(p^{(1)} - p^{(2)}\right)^{\n_3}+  \eta^{\n_2\n_3}\left(p^{(2)} - p^{(3)}\right)^{\n_1} +  \eta^{\n_1\n_3}\left(p^{(3)} - p^{(1)}\right)^{\n_2} \biggr\} \nn\\
&=& \kappa
\int \prod_{i=1}^3 dp^{(i)} \d\left( \sum_{i=1}^3 p^{(i)} \right) h_{\m_1\n_1}(p^{(1)})\,h_{\m_2\n_2}(p^{(2)})\,h_{\m_3\n_3}(p^{(3)})\nn\\
&& \biggl\{ \eta^{\m_1\m_2}p^{(1)\m_3} + \eta^{\m_2\m_3}p^{(2)\m_1} +  \eta^{\m_3\m_1}p^{(3)\m_2}\biggr\} 
\biggl\{ \eta^{\n_1\n_2}p^{(1)\n_3}+ \eta^{\n_2\n_3}p^{(2)\n_1} +  \eta^{\n_3\n_1}p^{(3)\n_2}  \biggr\}
\eeq 
where $\kappa =\frac{\sqrt{3} g}{2^{18}}=  \sqrt{32\pi G_{10}}$. This is precisely the three-graviton-scattering amplitude of the Einstein gravity \cite{TLeeEPJ2018,DeWitt1967,Schwarz1982,Sannan1986}. 

We may rewrite the cubic term in the non-Abelian gauge field action, which is obtained from the three open string scattering amplitude \cite{TLeeJKPS2017,TLee2017cov} as follows 
\beq \label{threegauge}
{\cal A}_{[3]{\rm gauge}} &=& \frac{g}{3} \int \prod_{i=1}^3 dp^{(i)} \d\left( \sum_{i=1}^3 p^{(i)} \right)  {\rm tr}~ \left(A_{\m_1}(p^{(1)})\,A_{\m_2}(p^{(2)})\,A_{\m_3}(p^{(3)})\right)\nn\\
&&
\biggl\{\eta^{\m_1\m_2}\left(p^{(1)} - p^{(2)}\right)^{\m_3} +\eta^{\m_2\m_3}\left(p^{(2)} - p^{(3)}\right)^{\m_1}
+\eta^{\m_3\m_1}\left(p^{(3)} - p^{(1)}\right)^{\m_2} \biggr\}  \nn\\
&=& g \int \prod_{i=1}^3 d p^{(i)} \d \left(\sum_{i=1}^3 p^{(i)} \right)\, p^{(1)\m} \text{tr} \left(A(1)^\n [ A(2)_\n, A_\m(3)] \right)   
\eeq 
where $g = \frac{2^6}{3^{\frac{9}{2}}} g_{\rm string}$. 
Comparing the three-gluon-scattering amplitude, Eq. (\ref{threegauge}) with the three-graviton-scattering amplitude, obtained from the cubic closed string field theory here Eq. (\ref{threegrav}), we can see the main point of the double copy theory in its simplest form: "gravity = gauge $\times$ gauge". 

\section{Conclusions and Discussions}

 The construction of BRST invariant covariant closed string field theory has been one of most outstanding problems in string theory. In this study, we directly extended the Witten's cubic open string field theory, and introduced double layers to open string world-sheets. The string world trajectory of open string can be 
mapped onto a unit disk on the complex $\o$-plane: The end points of the open strings form the boundary of the unit disk. The result showed that if we extend the range of 
spatial coordinate from $[0,\pi]$ for open string to $[-\pi, \pi]$ with periodic boundary condition for closed string, the world trajectory of string makes a closed curve in the complex $\o$-plane. Therefore, it is possible to describe the cubic closed 
string field theory, which is BRST invariant, extending the Witten's open string field theory. When we mapped the string world-sheet further on the complex $z$-plane, we confirmed that the extended string theory correctly describes closed string: The complex $z$-plane where the upper half only is covered by the open string is now 
fully covered and the complex $z$-plane is symmetric under reflection. It is not 
difficult to identify the overlapping condition in terms of the spatial string coordinate $\s$, which leads us to the extended string world-sheet of closed string. 
The three-closed-string vertex was constructed and found to respect the KLT structure: Their Neumann functions are completely factorized 
into those of corresponding open string. Finally we calculated explicitly the three 
graviton scattering amplitude, which is in perfect agreement with the Einstein gravity. 

We may extend this work along several directions: The first one is the background independent formulation of closed string field theory. As proposed in  \cite{Strominger1987nucl,Hata1986PLB,Horowitz1986PRL}, this cubic closed string field theory may be obtained from pure cubic field theory as we expand the pure cubic string field action around a classical solution. With the explicit expression of the $*$ operation between closed string fields, we may be able to make their arguments more concrete. It is also interesting to compare this cubic closed string field theory with closed string field theory \cite{Zwiebach1992nucl390,Hata1994,Zwiebach1998ann,Sen2016,Sen2017,Lacroix2017} based on the Batalin-Vilkovisky formulation \cite{Batalin1981,Batalin1983prd,Batalin1984nucl234,Vilkovisky1984,Batalin1985}. We need to clarify the relationship between these two approaches in future works.  

One of the important subjects in string theory is to construct open and closed string theory
\cite{Shapiro1987PLB,Kugo1998,Asakawa1998,Alisha2002,Takahashi2003,Gomis2004,Moosavian,Maccaferri2021}
 with a consistent coupling of open and closed strings. Adding Witten's cubic open string field theory action to the closed string field action constructed here with a proper coupling between the open and closed strings simplifies the construction of open-closed string field theory.
  We expect that the BRST invariance would play an important role again. Note that the vertex operator of three closed string, Eq. (\ref{vertexclosed3}) is completely factorized into those of three open string except for zero modes. This is a manifestation of the KLT relation \cite{Kawai1986} of the first quantized theory at the level of second quantized theory. If we are restricted to the spin two sector, this factorization may lead us to the recent works on the double copy theory \cite{BernPRL2010,Oxburgh2013}. This suggests that the open-closed string field theory may be the best framework to explore the essence of the double copy theory. 

Once we have constructed the closed string field theory, we may couple the closed 
string field to various $D$-branes, which play a role of sources for the closed string 
field. Solving the classical equation of motion for the closed string field, may leads us to $Dp$-brane classical solutions in terms of massless fields at large distance.  It implies that the double copy technique is also useful to study classical solutions of gravity theory 
\cite{Monteiro2014,Kim2020} and the closed 
string field theory may be the unified framework behind the double copy theory.

\vskip 1cm

\begin{acknowledgments}

This work was supported by the National Research Foundation of Korea(NRF) grant
funded by the Korea government (MSIT) (2021R1F1A106299311). 
\end{acknowledgments}

\begin{appendix}

\section{Calculations of the Neumann Functions of Closed Strings}

We may write the closed string Green's function as 
\beq \label{Greenfour}
G(\r_r, \rho^\prime_s) &=& \ln \vert z - \zp \vert \nn\\
&=& - \d_{rs} \left\{\sum_{n=1} \frac{1}{2n} \left(\omega^{-n}_+ \omega^{\prime n}_- 
+ \omega^{*-n}_+ \omega^{\prime * n}_-\right) - \max(\xi, \xi^\prime)  \right\} \nn\\
&& + \bar C^{rs}_{00} + \sum_{n=1} \left(\bar C^{rs}_{n0} \omega^n_r + \bar C^{rs}_{-n0} \omega^{*n}_r \right) + \sum_{m=1} \left(\bar C^{rs}_{0m} \omega^{\prime m}_s + \bar C^{rs}_{0-m} \omega^{\prime * m}_s \right) \nn\\
&& +\sum_{n, m \ge 1 } \Biggl\{ \bar C^{rs}_{nm}\omega^n_r \omega^{\prime m}_s + \bar C^{rs}_{-nm}
\omega^{ * n}_r \omega^{\prime  m}_s + \bar C^{rs}_{n-m}\omega^n_r \omega^{\prime * m}_s + 
\bar C^{rs}_{-n-m}\omega^{ * n}_r \omega^{* \prime m}_s\Biggr\}
\eeq
where $\o_r = e^{\zeta_r} = e^{\xi_r+ i\eta_r}, ~~~\op_s = e^{\xi^\prime_r + i\eta^{\prime}_s}$, and
\beq
(\o_+,\o_-) &=&  \left\{ 
\begin{array}{ll}
(\o_r, \op_s) , & ~~\mbox{for} ~~\xi_r \ge \xi^\prime_s   \\ 
(\op_s, \o_r) , & ~~\mbox{for} ~~\xi_r \le \xi^\prime_s 
\end{array} \right . 
\eeq
Here we shall calculate the Neumann functions of closed string theory in detail
\begin{itemize}
\item $\bar C^{rs}_{00}$, for $r \not= s$, \\
Taking $\zp \rightarrow Z_s$ ($\zeta^\prime_s \rightarrow - \infty$), and $z \rightarrow Z_r$,
we have 
\beq
\bar C^{rs}_{00} = \ln \vert Z_r -Z_s \vert . 
\eeq 

\item $\bar C^{rs}_{00}$, for $r = s$, \\
If we take the limit $z^\prime_r \rightarrow Z_r$, 
\beq
G(\rho_r, \rho^\prime_s) &=& \xi_r + \sum_n \bar C^{rs}_{n0} e^{|n| \xi_r} e^{in\eta_r} = \ln \vert z-Z_r \vert .
\eeq 
Taking the limit $z_r \rightarrow Z_r$, we find 
\beq \label{c0058}
\bar C^{rr}_{00} &=& \ln |z_r-Z_r| - \xi_r. 
\eeq 
Since near $z_r = Z_r$, we expand $e^{-\z_r}$ 
\beq
e^{-\z_r} &=& \frac{a_r}{(z_r-Z_r)} + \sum_{n=0} c^{(r)}_n (z_r-Z_r)^n .
\eeq 
It follows that in the limit $z_r \rightarrow Z_r$
\beq \label{limit61}
\xi_r  = \ln |z_r - Z_r| - \ln a_r .
\eeq
Using  Eq. (\ref{c0058})  and Eq.(\ref{limit61}) in the limit, where $z_r \rightarrow Z_r$, we obtain
\beq
\bar C^{rr}_{00} &=& \ln a_r . 
\eeq 

\item $\bar C^{rs}_{n0}, ~ n\not= 0$: Differentiating Eq. (\ref{Greenfour}) with respect to $\zeta_r$, 
(assuming $\xi_r \ge \xi^\prime_s$)
\beq
\frac{\p}{\p \zeta_r} G(\rho_r, \rho^\prime_s) &=& \frac{1}{2} \left(\frac{\p z}{\p \zeta_r} \right)
\frac{1}{z-\zp} \nn\\
&=& \omega_r \frac{\p}{\p \omega_r} G(\rho_r, \rho^\prime_s) \nn\\
&=& \d_{rs} \left\{\frac{1}{2} \sum_{n \ge 1} \o^{-n}_r \o^{\prime n}_s  +\half 
\right\}+ \sum_{n=1} n \bar C^{rs}_{n0} \omega^n_r  \nn\\
&& + \sum_{n,m \ge 1}  \Bigl\{n \bar C^{rs}_{nm} \o^n_r \o^{\prime m}_s + 
n \bar C^{rs}_{n-m} \o^n_r \o^{\prime * m}_s \Bigr\} \label{Diff1}
\eeq
where we make use of
\beq
\frac{\p}{\p \zeta_r} = \o_r \frac{\p}{\p \o_r} = \frac{1}{2} \left(\frac{\p}{\p \xi_r} -i \frac{\p}{\p \eta_r} \right) .
\eeq
Taking the limit, $z^\prime_s \rightarrow Z_s$ $(\o^\prime_s \rightarrow 0)$, of Eq. (\ref{Diff1}) 
\beq
\d_{rs} + 2\sum_{n \ge 1} n \bar C^{rs}_{n0} \o^n_r = \left(\frac{\p z}{\p \zeta_r} \right)
\frac{1}{z-Z_s} .
\eeq
By performing the contour integral around $\o_r = 0~ (z_r = Z_r)$, 
\beq
\oint_{\o_r = 0} d\o_r\, \o^{-n-1}_r \left(\d_{rs} + 2\sum_{m \ge 1} \bar C^{rs}_{m0} \o^m_r \right) = 
\oint_{\o_r = 0} d\o_r\, \o^{-n-1}_r  \left(\frac{\p z}{\p \zeta_r} \right) \frac{1}{z-Z_s} 
\eeq
we get 
\beq
\bar C^{rs}_{n0} &=& \bar C^{sr}_{0n} = \frac{1}{2n} \oint_{Z_r} \frac{d z}{2\pi i} \frac{1}{z-Z_s} e^{-n\zeta_r(z)}, ~~~n \ge 1, \label{Cn0}
\eeq 
Here we used $d\o_r = \o_r d \zeta_r$ and
\beq
\frac{d \o_r}{\o_r} \frac{\p z}{\p \z_r} = d \z_r \frac{\p z}{\p \z_r} = dz .
\eeq 

$\bar C^{rs}_{-n 0}$: From reality condition of the Green's function
\beq
G (\rho_r, \rho^\prime_s) = G (\rho_r, \rho^\prime_s)^* ,
\eeq 
we have 
\beq
\bar C^{rs}_{nm} &=& \bar C^{rs *}_{-n -m} . 
\eeq

\item $\bar C^{rs}_{nm}$: Differentiating Eq. (\ref{Diff1}) with respect to $\zeta^\prime_s$, 
\beq \label{Diff2}
\frac{\p }{\p \zeta_r}\frac{\p }{\p \zeta^\prime_s} G(\r_r, \rho^\prime_s) 
&=& \frac{1}{2} \left(\frac{\p z}{\p \zeta_r} \right) \left(\frac{\p \zp}{\p \zeta^\prime_s}\right)
\frac{1}{(z-\zp)^2} \nn\\
&=& \d_{rs} \frac{1}{2} \sum_{n \ge 1} n \o^{-n}_r \o^{\prime n}_s + \sum_{n, m \ge 1} nm
\bar C^{rs}_{nm} \o^n_r \o^{\prime m}_s .
\eeq
Then performing the contour integral around $\op_s = 0$ $(z^\prime_s = Z_s)$, 
$\oint d\o_r \oint d\op_s \o^{-n-1}_r \o^{\prime -m -1}_s$ of Eq.~(\ref{Diff2}), 
we get
\beq
\bar C^{rs}_{nm} &=& \frac{1}{2nm} \oint_{Z_r} \frac{dz}{2\pi i} \oint_{Z_s} \frac{d \zp}{2\pi i} 
\frac{1}{(z-\zp)^2} e^{-n\zeta_r(z) - m \zeta^\prime_s(\zp)}, ~~~ n, m \ge 1 
\eeq

\item $\bar C^{rs}_{-n0}$: Differentiate Eq. (\ref{Greenfour}) with respect to $\zeta_r^*$, we have 
\beq
\frac{\p}{\p \zeta^*_r} G_C(\rho_r, \rho^\prime_s) &=& \frac{1}{2} \left(\frac{\p z^*}{\p \zeta^*_r} \right)
\frac{1}{z^*-\zp^*} \nn\\
&=& \omega^*_r \frac{\p}{\p \omega^*_r} G_C(\rho_r, \rho^\prime_s) \nn\\
&=& \d_{rs} \left\{\frac{1}{2} \sum_{n \ge 1} \o^{*-n}_r \o^{\prime * n}_s  +\half 
\right\}+ \sum_{n=1} n \bar C^{rs}_{-n0} \omega^{* n}_r  \nn\\
&& + \sum_{n,m \ge 1}  \Bigl\{n \bar C^{rs}_{-nm} \o^{*n}_r \o^{\prime  m}_s + 
n \bar C^{rs}_{-n-m} \o^{*n}_r \o^{\prime * m}_s \Bigr\} \label{Diff*}
\eeq
where we use
\beq
\frac{\p}{\p \zeta^*_r} = \o^*_r \frac{\p}{\p \o^*_r} = \frac{1}{2} \left(\frac{\p}{\p \xi_r} +i \frac{\p}{\p \eta_r} \right) .
\eeq
Then we take the limit, $\zp^*_s \rightarrow Z^*_s$ $(\o^{\prime *}_s \rightarrow 0)$, 
\beq
\d_{rs} + 2\sum_{n \ge 1} n \bar C^{rs}_{-n0} \o^{* n}_r = \left(\frac{\p z^*}{\p \zeta^*_r} \right)
\frac{1}{z^*-Z^*_s} .
\eeq
From performing the contour integral around $\o^*_r = 0~ (z^*_r = Z^*_r)$, 
\beq
\oint_{\o^*_r = 0} d\o^*_r\, \o^{* -n-1}_r \left(\d_{rs} + 2\sum_{m \ge 1} \bar C^{rs}_{-m0} \o^{* m}_r \right) = 
\oint_{\o^*_r = 0} d\o^*_r\, \o^{*-n-1}_r  \left(\frac{\p z^*}{\p \zeta^*_r} \right) \frac{1}{z^*-Z^*_s} 
\eeq
it follows that  
\beq
\bar C^{rs}_{-n0} &=& \bar C^{sr}_{0-n} =- \frac{1}{2n} \oint_{Z^*_r} \frac{d z^*}{2\pi i} \frac{1}{z^*-Z^*_s} e^{-n\zeta^*_r(z^*)}, ~~~n \ge 1, \label{C-n0}
\eeq 
Here we used $d\o^*_r = \o^*_r d \zeta^*_r$ and
\beq
\frac{d \o^*_r}{\o^*_r} \frac{\p z^*}{\p \z^*_r} = d \z^*_r \frac{\p z^*}{\p \z^*_r} = dz^* .
\eeq 

\item $\bar C^{rs}_{-nm}$: Differentiate Eq. (\ref{Diff*}) with respect to $\zeta_s^\prime$,  
\beq
\frac{\p}{\p \zeta^*_r}\frac{\p}{\p \zeta^\prime_s} G(\rho_r, \rho^\prime_s) = \sum_{n, m \ge 1} n m \bar C^{rs}_{-n m} \omega^{* n}_r \omega^{\prime *}_s = 0.
\eeq 
It implies 
\beq
\bar C^{rs}_{-n m} = 0, ~~~ n, m \ge 1 .  
\eeq 

\item $\bar C^{rs}_{-nm}$: Differentiate Eq. (\ref{Diff*}) with respect to $\zeta_s^{\prime *}$,    
\beq \label{Diff2*}
\frac{\p }{\p \zeta^*_r}\frac{\p }{\p \zeta^{\prime *}_s} G_C(\r_r, \rho^\prime_s) &=& 
\frac{1}{2} \left(\frac{\p z^*}{\p \zeta^*_r} \right) \left(\frac{\p \zp^*}{\p \zeta^{\prime *}_s}\right)
\frac{\p}{\p \zp^*} \left(\frac{1}{z^*-\zp^*} \right) \nn\\
&=& \frac{1}{2} \left(\frac{\p z^*}{\p \zeta^*_r} \right) \left(\frac{\p \zp^*}{\p \zeta^{\prime *}_s}\right)
\frac{1}{(z^*-\zp^*)^2} \nn\\
&=& \d_{rs} \frac{1}{2} \sum_{n \ge 1} n \o^{* -n}_r \o^{\prime * n}_s + \sum_{n, m \ge 1} nm
\bar C^{rs}_{-n-m} \o^{* n}_r \o^{\prime * m}_s .
\eeq
Then, performing the contour integral around $\op{}^*_s = 0$ $(z^{\prime *}_s= Z^*_s)$, 
$\oint d\o^*_r \oint d\op{}^*_s \o^{*-n-1}_r \o^{\prime *-m -1}_s$ of Eq.~(\ref{Diff2*}), 
we get
\beq
\bar C^{rs}_{-n-m} &=& \frac{1}{2nm} \oint_{Z_r} \frac{dz^*}{2\pi i} \oint_{Z_s} \frac{d \zp{}*}{2\pi i} 
\frac{1}{(z^*-\zp{}^*)^2} e^{-n\zeta^*_r(z^*) - m \zeta^{\prime *}_s(\zp{}^*)}, ~~~ n, m \ge 1 . 
\eeq 

If we Choose $Z_r, ~ r = 1, 2, \dots, N$ be on the real line, 
\begin{subequations}
\beq
\bar C^{rs}_{n0} &=& \bar C^{rs}_{-n0} = \half \bar N^{rs}_{n0} = \frac{1}{2n} \oint_{Z_r} \frac{d z}{2\pi i} \frac{1}{z-Z_s} e^{-n\zeta_r(z)}, ~~~n \ge 1,, \\
\bar C^{rs}_{nm} &=& \bar C^{rs}_{-n-m} = \half \bar N^{rs}_{mn} =\frac{1}{2nm} \oint_{Z_r} \frac{dz}{2\pi i} \oint_{Z_s} \frac{d \zp}{2\pi i} 
\frac{1}{(z-\zp)^2} e^{-n\zeta_r(z) - m \zeta^\prime_s(\zp)}, ~~~ n, m \ge 1 , \\
\bar C^{rs}_{n-m} &=& \bar C^{rs}_{-n m} = 0. 
\eeq 
\end{subequations}

\end{itemize}

The Fock space representation of the vertex operator may be written as
\beq
\vert {V}{[N]} \rangle &=& \exp\left(-\sum_r \xi_r \left(L^{(r)}_0 + \tilde  L^{(r)}_0 \right)\right)
e^{ \sum_r \frac{\xi_r}{2} \left( \left(p^r_L\right)^2 + \left(p^r_R\right)^2 \right)} \nn\\
&&
\exp \Biggl\{\frac{1}{4} \sum_{r,s} \sum_{n, m \ge 1} \Biggl(\bar C^{rs}_{nm} e^{n \xi_r + m \xi_s}\, (\a^{(r)}_n + \tilde\a^{(r)\dag}_n ) \cdot (\a^{(s)}_m + \tilde\a^{(s)\dag}_m )\nn\\
&&+\bar C^{rs}_{-n-m} e^{n \xi_r + m \xi_s}\, (\a^{(r)\dag}_n + \tilde\a^{(r)}_n ) \cdot (\a^{(r)\dag}_m + \tilde\a^{(r)}_m ) \Biggl) + \frac{1}{4} \sum_{r,s} \Biggr(\sum_{n \ge 1} \Bigr(
\bar C^{rs}_{n0} e^{n\xi_r} \left(\a^{(r)}_n + \tilde\a^{(r)\dag}_n \right) \cdot P_0^{(s)}\nn\\
&& + \bar C^{rs}_{-n0}e^{n\xi_r} \left(\a^{(r)\dag}_n + \tilde\a^{(r)}_n \right) \cdot P_0^{(s)} \Bigr)
+ \sum_{m \ge 1} P_0^{(r)} \cdot \Bigl(\bar C^{rs}_{0m} e^{m\xi_s}\left(\a^{(s)}_m + \tilde \a^{(s)\dag}_m \right) \nn\\
&& +\bar C^{rs}_{0-m} e^{m\xi_s} \left(\a^{(s)\dag}_m + \tilde \a^{(s)}_m\right)\Bigr)  + \bar C^{rs}_{00} P_0^{(r)} \cdot P_0^{(s)} \Biggr)\Biggr\} \vert 0 \rangle\nn\\
&=& e^{-\sum_r 2 \bar C^{rr}_{00}} 
\exp \Biggl\{\frac{1}{4} \sum_{r,s} \sum_{n, m \ge 1} \Biggl(\bar C^{rs}_{nm}\,\tilde\a^{(r)\dag}_n \cdot \tilde\a^{(s)\dag}_m +\bar C^{rs}_{-n-m} \, \a^{(r)\dag}_n \cdot \a^{(r)\dag}_m \Biggl) + \frac{1}{4} \sum_{r,s} \Biggr(\sum_{n \ge 1} \Bigr(
\bar C^{rs}_{n0}\tilde\a^{(r)\dag}_n\cdot P_0^{(s)}\nn\\
&& + \bar C^{rs}_{-n0} \a^{(r)\dag}_n\cdot P_0^{(s)} \Bigr)
+ \sum_{m \ge 1} P_0^{(r)} \cdot \Bigl(\bar C^{rs}_{0m}\tilde \a^{(s)\dag}_m +
\bar C^{rs}_{0-m} \a^{(s)\dag}_m \Bigr) + \bar C^{rs}_{00} P_0^{(r)} \cdot P_0^{(s)} 
\Biggr)\Biggr\} \vert 0 \rangle  .
\eeq 
Note that 
\beq
\exp\left(-\sum_r \xi_r \left(L^{(r)}_0 + \tilde  L^{(r)}_0 \right)\right)
e^{ \sum_r \frac{\xi_r}{2} \left( \left(p^r_L\right)^2 + \left(p^r_R\right)^2 \right)} &=& 
e^{\sum_r 2 \xi_r} \exp \left(-\sum_r \xi_r (\a^{(r)\dag} \cdot \a^{(r)}
+ \tilde\a^{(r)\dag} \cdot \tilde \a^{(r)})\right) \nn\\
\exp\left\{\sum_r 2\xi_r + \sum_{r,s} \bar C^{rs}_{00} p^{(r)} \cdot p^{(s)} \right\} &=& 
\exp\left\{2\sum_r \bar C^{rr}_{00} \left(\frac{(p^{(r)})^2}{2} -1\right) \right\} \prod_{r <s} \vert Z_r - Z_s \vert^{2 p^{(r)} \cdot p^{(s)}} 
\eeq 
where we make use of 
\beq
\xi_r = - \bar C^{rr}_{00} + \ln |z_r - Z_r| = - \bar C^{rr}_{00} + \ln |\e|, ~~\text{and}~~
\bar C^{rs}_{00} =  \ln \vert Z_r -Z_s \vert .
\eeq 

To summarize, we may write for $n \ge 1$, and $m \ge 1$, 
\begin{subequations}
\beq
\bar C^{rs}_{00} &=& \bar N^{rs}_{00} = \ln \vert Z_r -Z_s \vert ,~~~ r\not=s, \label{a33a}\\
\bar C^{rr}_{00} &=& \bar N^{rr}_{00} = \ln |a_r|, ~~~ a_r = 2 ~~\text{for}~N=2 \label{a33b}\\
\bar C^{rs}_{n0} &=& \bar C^{rs}_{-n0} = \half \bar N^{rs}_{n0} = \frac{1}{2n} \oint_{Z_r} \frac{d z}{2\pi i} \frac{1}{z-Z_s} e^{-n\zeta_r(z)}, ~~~n \ge 1, \label{a33c}\\
\bar C^{rs}_{nm} &=& \bar C^{rs}_{-n-m} = \half \bar N^{rs}_{mn} =\frac{1}{2nm} \oint_{Z_r} \frac{dz}{2\pi i} \oint_{Z_s} \frac{d \zp}{2\pi i} 
\frac{1}{(z-\zp)^2} e^{-n\zeta_r(z) - m \zeta^\prime_s(\zp)}, ~~~ n, m \ge 1 , \label{a33d}\\
\bar C^{rs}_{n-m} &=& \bar C^{rs}_{-n m} = 0.
\eeq
\end{subequations}
For $N \ge 2$, we find that the $N$-vertex operator may be expressed as 
\beq \label{VNexp}
\vert V{[N]} \rangle &=&
\exp \Biggl\{ 2\sum_r \ln|a_r| \left( \frac{\left(p^{(r)}\right)^2}{2} -1 \right) 
\Biggr\} \prod_{r <s} \vert Z_r - Z_s \vert^{2 p^{(r)} \cdot p^{(s)}}  \nn\\
&&
\exp  \Biggl\{
\sum_{r,s} \Bigl( \sum_{n, m \ge 1} \frac{1}{2} \bar N^{rs}_{nm}\,\frac{\a^{(r)\dag}_n}{2} \cdot 
\frac{\a^{(r)\dag}_m}{2} 
+ \sum_{n \ge 1}\bar N^{rs}_{n0} \frac{\a^{(r)\dag}_n}{2} \cdot \frac{p^{(s)}}{2} \Bigr) 
\Biggr\} \nn\\
&&  
\exp  \Biggl\{
\sum_{r,s} \Bigl( \sum_{n, m \ge 1} \frac{1}{2} \bar N^{rs}_{nm}\,\frac{\tilde\a^{(r)\dag}_n}{2} \cdot \frac{\tilde\a^{(r)\dag}_m}{2} 
+ \sum_{n \ge 1}\bar N^{rs}_{n0} \frac{\tilde\a^{(r)\dag}_n}{2} \cdot \frac{p^{(s)}}{2} \Bigr) \Biggr\}
\eeq 
Note that the Kawai-Lewellen-Tye (KLT) relations are manifest at the second quantized level.

\end{appendix}

%\begin{references}

%\begin{figure}[htbp]
%   \begin {center}
   %\input epsf.tex
%    \epsfxsize=0.7\hsize
%
% Specify the picture file name to be included.
%	\epsfbox{circle.pdf}
%   \end {center}
%   \caption {\label{circle} The magic circles and the critical circle}
%\end{figure}

%\begin{figure}[htbp]
%   \begin {center}
   %\input epsf.tex
%    \epsfxsize=0.7\hsize
%
% Specify the picture file name to be included.
%	\epsfbox{quonfig.pdf}
%   \end {center}
%   \caption {\label{quon} Q-deformed statistics}
%\end{figure}

%\end{references}

\end{document}